\documentclass[namedreferences]{solarphysics}
\usepackage[hyperref,optionalrh,solaromanenum]{spr-sola-addons} 
\usepackage{graphicx}                    
\usepackage{color}                       
\usepackage{breakurl}                         

\begin{document}

\begin{article}

\begin{opening}

\title{Analysis of the flux growth rate in emerging active regions on the Sun}

%
\author[addressref={aff1,aff2}]{\inits{V.I.}\fnm{V.I.}~\lnm{Abramenko}}
\author[addressref=aff1, corref, email={alex.s.kutsenko@gmail.com}]{\inits{A.S.}\fnm{A.S.}~\lnm{Kutsenko}}
\author[addressref=aff1]{\inits{O.I.}\fnm{O.I.}~\lnm{Tikhonova}}
\author[addressref={aff3,aff4}]{\inits{V.B.}\fnm{V.B.}~\lnm{Yurchyshyn}}

\address[id=aff1]{Crimean Astrophysical Observatory, Russian Academy of Science, Nauchny, Bakhchisaray, 298409, Crimea}
\address[id=aff2]{Central (Pulkovo) Astronomical Observatory, Russian Academy of Science (GAO RAN), Pulkovskoye ch. 65,  Saint-Petersburg, 196140, Russia}
\address[id=aff3]{Big Bear Solar Observatory, New Jersey Institute of Technology, Big Bear City, CA 92314, USA}
\address[id=aff4]{Korea Astronomy and Space Science Institute, 776 Daedeok-daero, Yuseong-gu, Daejeon, 305-348, South Korea}

%
\runningauthor{V.I. Abramenko \emph{et al.}}
\runningtitle{Analysis of the flux growth rate in emerging ARs on the Sun}

\begin{abstract}

We studied the emergence process of 42 active region (ARs) by analyzing the time derivative, $R(t)$, of the total unsigned flux. Line-of-sight magnetograms acquired by the \textit{Helioseismic and Magnetic Imager} (HMI) onboard the \textit{Solar Dynamics Observatory} (SDO) were used. A continuous piecewise linear fitting to the $R(t)$-profile was applied to detect an interval, $\Delta t\textsubscript{2}$, of nearly-constant  $R(t)$ covering one or several local maxima. The averaged over $\Delta t\textsubscript{2}$ magnitude of $R(t)$ was accepted as an estimate of the maximal value of the flux growth rate, $R\textsubscript{MAX}$, which varies in a range of (0.5-5)$\times$10\textsuperscript{20} Mx hour\textsuperscript{-1} for active regions with the maximal total unsigned flux of (0.5-3)$\times$10\textsuperscript{22} Mx. The normalized flux growth rate, $R\textsubscript{N}$, was defined under an assumption that the saturated total unsigned flux, $F\textsubscript{MAX}$, equals unity. Out of 42 ARs in our initial list, 36 event were successfully fitted and they form two subsets (with a small overlap of 8 events): the ARs with a short (\textless13 hours) interval $\Delta t\textsubscript{2}$ and a high (\textgreater0.024 hour\textsuperscript{-1}) normalized flux emergence rate, $R\textsubscript{N}$, form the "rapid" emergence event subset. The second subset consists of "gradual" emergence events and it is characterized by a long (\textgreater13 hours) interval $\Delta t\textsubscript{2}$ and a low $R\textsubscript{N}$ (\textless0.024 hour\textsuperscript{-1}). In diagrams of $R\textsubscript{MAX}$ plotted versus $F\textsubscript{MAX}$, the events from different subsets are not overlapped and each subset displays an individual power law. The power law index derived from the entire ensemble of 36 events is 0.69$\pm$0.10. The "rapid" emergence is consistent with a "two-step" emergence process of a single twisted flux tube. The  "gradual" emergence is possibly related to a consecutive rising of several flux tubes emerging at nearly the same location in the photosphere.

\end{abstract}

%

\end{opening}

%
 \section{Introduction}
	 \label{sec-Introduction} 
	 
Emergence of new active regions (ARs) is believed to be manifestation of the solar dynamo action deep in the convective zone \citep{Babcock1961, Leighton1969}. \cite{Parker1975} suggested that a magnetic $\Omega$-shaped flux tube emerges due to magnetic buoyancy under the resistance of aerodynamic drag. The process is not available for direct observations, so methods of numerical simulations are still the main source of information on the details of the emergence process.

Significant progress was made since the first two-dimensional magnetohydrodynamic (MHD) simulations of flux emergence performed by \cite{Schuessler1979}. Thus, \cite{Spruit1981} suggested a thin-flux-tube approximation, and \cite{MorenoInsertis1996} and \cite{Emonet1998} performed compressible MHD simulations for a twisted flux tube. Flux emergence in the turbulent convective zone (CZ) was also simulated in the framework of the anelastic approximation (\textit{e.g.},  \cite{Jouve2009} and references therein). However, according to \cite{Fan2009}, the approximation does not work for the upper CZ (above 0.95 solar radius) due to large  gradients of physical parameters. The upper 20 Mm of the CZ thus became  the focus of intense research efforts.

Thus, \cite{Archontis2004} proposed a “two-step-emergence” model where a twisted flux rope, driven by buoyancy emerges from the depth of -20 Mm and slows down as it reaches the photosphere because the unmagnetized plasma becomes trapped between the rising flux and the photosphere hindering further emergence. During the intermediate phase, slow rise of the magnetized plasma is accompanied by enhanced expansion in the transverse direction. The second step of emergence is a runway expansion into the corona due to buoyant instability (see Figures 5 and 7 in \cite{Archontis2004}). The model was later elaborated in the framework of 2D and 3D compressible MHD simulations by \cite{Toriumi2010, Toriumi2011}.

\cite{Cheung2007} and \cite{Cheung2008} explored the flux emergence process through the top layers of the CZ and the photosphere using 3D radiative MHD simulations and taking into account the effect of compressibility, energy exchange via radiative transfer, and partial ionization. Only ARs with the total magnetic flux less than 10\textsuperscript{20} Mx (a typical flux of an ephemeral region \citep{Hagenaar2001}) were considered. The comparison of the simulation results with \textit{Hinode} observations \citep{Cheung2008} demonstrated the success of the simulations. Counterparts of observed high-speed down-flows, bright points, anomalous darkenings, transient kilogauss horizontal fields were reproduced by the model and explained.  Moreover, \cite{Cheung2008} have shown that the rate of emergence (“the rate of transport of the longitudinal flux”) is higher for the more twisted flux tubes. The undulating behavior of the field lines connecting the poles of a magnetic dipole (serpentine field lines) was demonstrated in the simulations as well. Authors argue that convective dynamics of photospheric granules is responsible for this effect that is ubiquitous in observations of flux emergence. In \cite{Cheung2010} the serpentine pattern as well as the lateral expansion of magnetic flux is demonstrated in case of emergence of a semi-torus. The process of flux emergence, sunspots formation, and subsequent decay was recently modeled by \cite{Rempel2014} starting from the rise of a semi-torus shaped magnetic flux of 1.7$\cdot$10\textsuperscript{22} Mx from the depth of 15.5 Mm. Authors have shown that large-scale subsurface flows can affect the resulting emerged magnetic structure.

The above brief discussion of the most relevant theoretical works is by no means a comprehensive review of numerical simulation efforts aimed to model the flux emergence on the Sun. For an extended review, we recommend the \textit{Living Review in Solar Physics} by \cite{Cheung2014}.

Impressive achievements of numerical simulations should be compared with observations and explain observed features, as they do frequently. However, for the further progress, efforts in opposite direction are desirable, \textit{i.e.}, characteristics of flux emergence that follow from observations, should complement the modeling progress.

When analyzing flux emergence using observations, one of the essential parameters to study  is the total unsigned flux and its time variations. Frequently the time derivative of the total unsigned flux, the flux emergence rate or flux growth rate, is explored as well.

\cite{Otsuji2011} studied 101 flux emergence events ranging from small ephemeral regions to large ARs by using  \textit{Hinode}/SOT filtergrams and magnetograms. They calculated the flux growth rate as the ratio of the saturated unsigned flux to the total emergence time and considered this value as the average flux growth rate, $<dF/dt> = F\textsubscript{MAX}/T$.  A relationship between $<dF/dt>$ and $F\textsubscript{MAX}$ was found: $<dF/dt> \propto F\textsubscript{MAX}\textsuperscript{0.57}$. Under a set of simplifications it was shown that the index should be 0.5,  which is close to 0.57. Authors consider this inference as a partial confirmation of an elaborated physical view of the solar flux emergence.

\cite{Centeno2012} studied the early emergence of two moderate-size ARs using SDO/HMI vector magnetic field data and Doppler velocities focusing on moving dipolar features observed between the leading and following parts of an AR. Their observed characteristics can be well explained in the framework of \cite{Cheung2010} modeling that predicts unavoidable appearance of serpentine-like field lines responsible for near-surface reconnection and the discharge of mass from rising magnetic flux tubes.

\cite{Khlystova2013} studied emergence during the first $\Delta t=12$ hours since the first appearance of a new flux using SOHO/MDI magnetograms and Dopplergrams of 224 emerging ARs. The flux emergence rate, $dF/dt$, was derived as the ratio of the maximum (during $\Delta t$) unsigned magnetic flux, $F\textsubscript{MAX}$, and the interval length, $\Delta t$.   Their scatter plot  (Figure 8 in \cite{Khlystova2013}) allows to infer a power law relationship between the $F\textsubscript{MAX}$ and $dF/dt$ with an index of about 0.4. Author explained the weak dependence between the parameters suggesting  that ARs emerge slowly and gradually fragment after fragment during a rather long time period (frequently up to several days), and the flux growth rate derived from the first half-day of emergence is not a comprehensive characteristic of an AR as a whole.

\cite{Toriumi2014} utilized SDO/HMI data for 23 emerging ARs to study horizontal divergent flows that are expected to accompany the emergence process according to the “two-step” emergence scenario \citep{Archontis2004}. In 13 cases, these flows were detected 30-100 minutes prior to the emergence onset. Authors suggested that the observed consecutive peaks are likely to indicate bifurcation of the emerging magnetic flux, which may consist of separate flux bundles \citep{Zwaan1985}. However, these  authors did not remove the well known artificial 24-hour oscillations in the magnetic flux (see, \textit{e.g.}, \cite{Liu2012, Smirnova2013, Kutsenko2016}) so that the resulting time profiles of the flux growth rate are contaminated by artificial strong peaks separated by 24-hour intervals (see Figures 1 and 7 in \cite{Toriumi2014}) making the growth rate measurements uncertain. Nevertheless, the \cite{Toriumi2014} suggestion is promising, and it should be investigated using improved and more accurate detection routines.

The magnetic flux emergence process is usually accompanied by interaction between emerging and pre-existing fluxes. This topic was scrutinized by \cite{Fu2016} who explored the problem in both theoretical and observational way. As a part of their investigation, they determined the emergence time for 116 flux emergence cases. Authors considered only the total unsigned flux time variations and did not calculate the flux growth rate. A two-step emergence behavior, when a gradual  emergence stage is followed by rapid emergence, was detected in 54 cases, whereas 46 cases were consistent with a single-phase (rapid) emergence process.

In the present study, we focus on the analysis of the time variations of the flux growth rate. Using SDO/HMI line-of-sight magnetograms, we derive this parameter as a time derivative of the total unsigned flux, which was preliminary filtered to remove the artificial 24-hour oscillations \citep{Kutsenko2016}. Our aim is to reveal how the flux growth rate varies during the rising phase in various active regions and to make an attempt to infer some information about the emerging structure basing on the flux growth rate profile.


 \section{Data Selection and Reduction}
	 \label{sec-DataSel}

\begin{figure}    
  	\centerline{\hspace*{0.015\textwidth}
  		\includegraphics[width=0.515\textwidth,clip=]{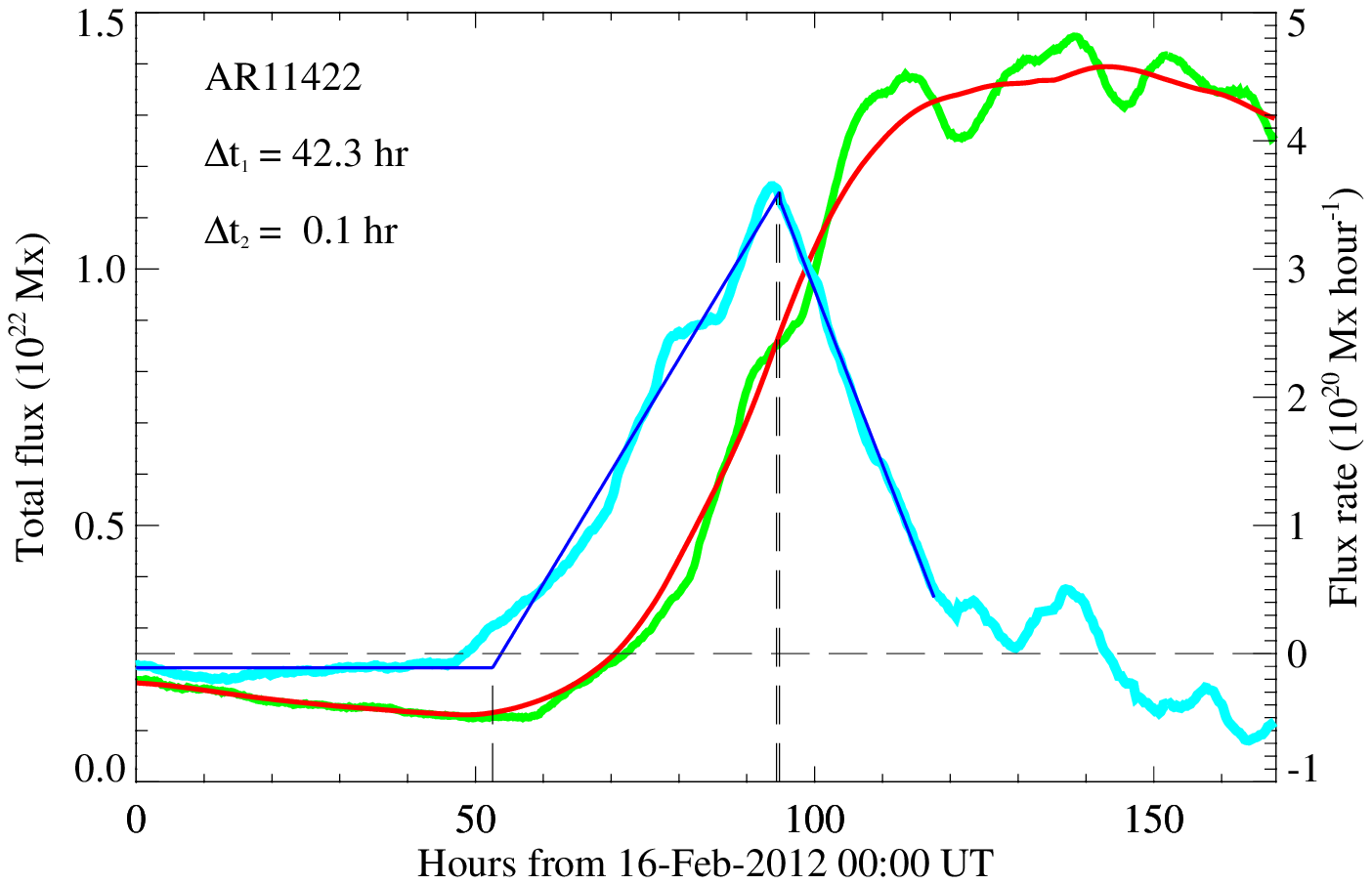}
  		\hspace*{-0.03\textwidth}
  		\includegraphics[width=0.515\textwidth,clip=]{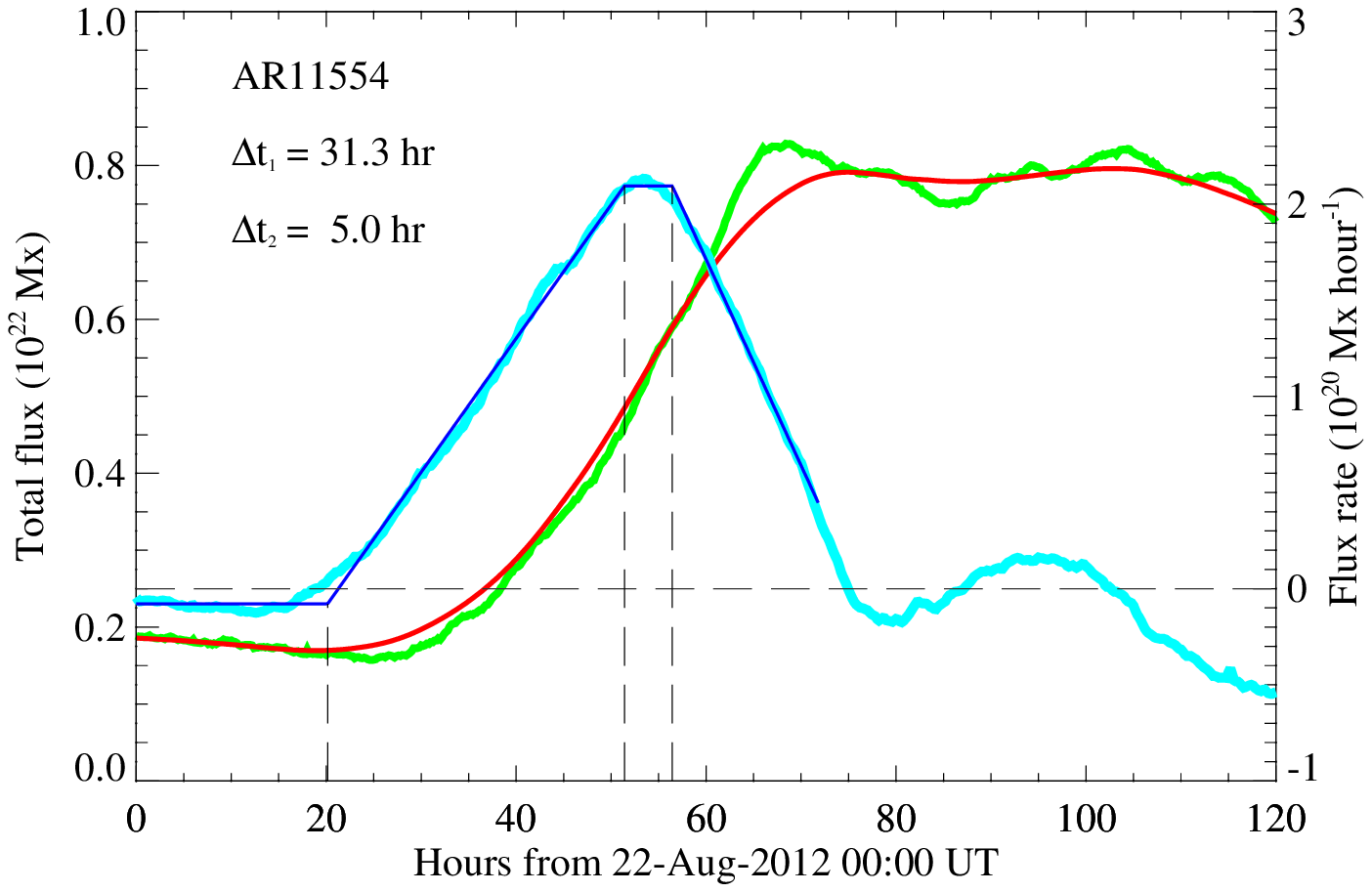}
  	}
  	\vspace{-0.35\textwidth}   
  	\centerline{\Large \bf     
  		\hspace{0.0 \textwidth}  \color{white}{(a)}
  		\hspace{0.415\textwidth}  \color{white}{(b)}
  		\hfill}
  	\vspace{0.31\textwidth}    
  	\centerline{\hspace*{0.015\textwidth}
  		\includegraphics[width=0.515\textwidth,clip=]{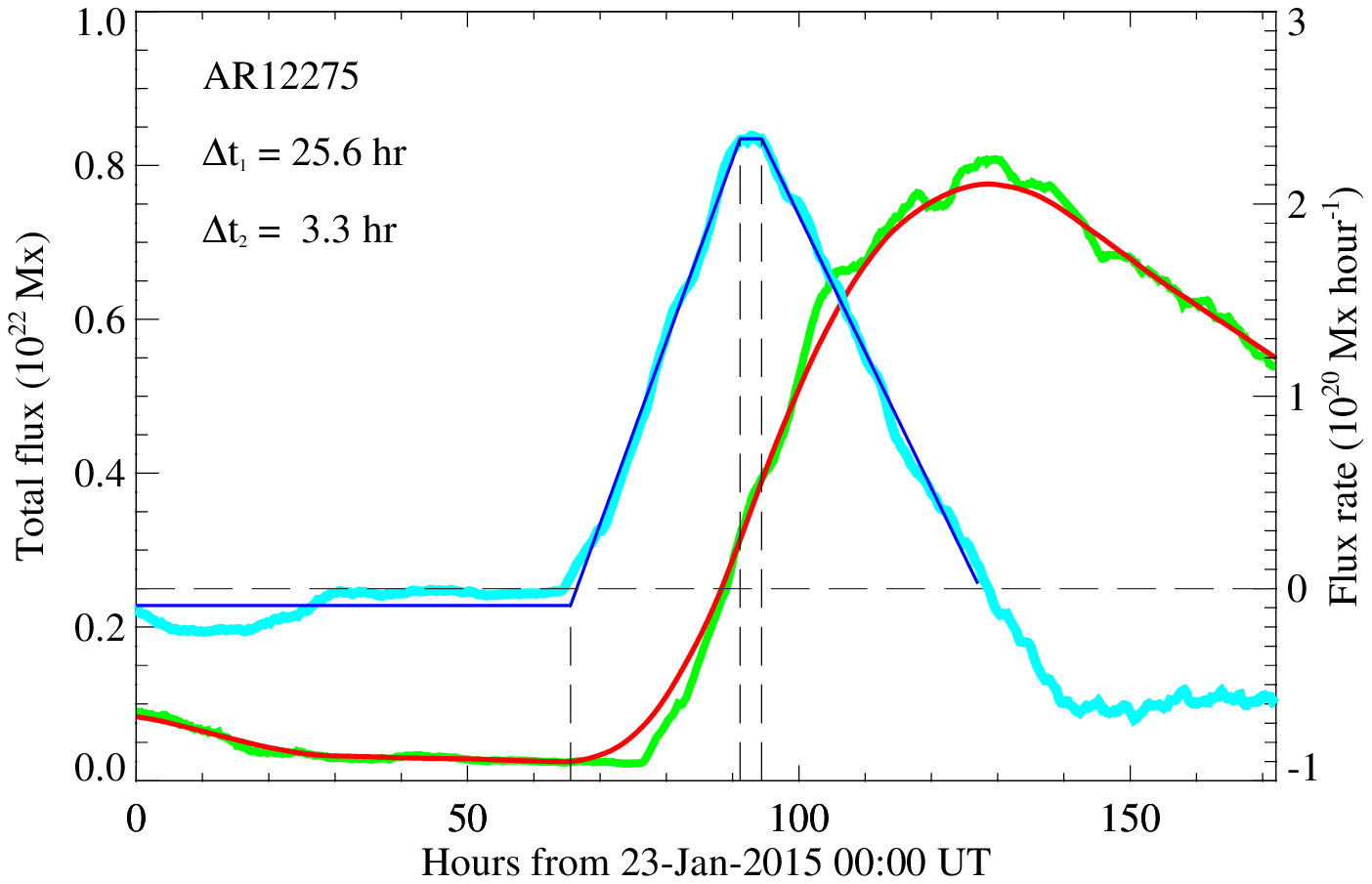}
  		\hspace*{-0.03\textwidth}
  		\includegraphics[width=0.515\textwidth,clip=]{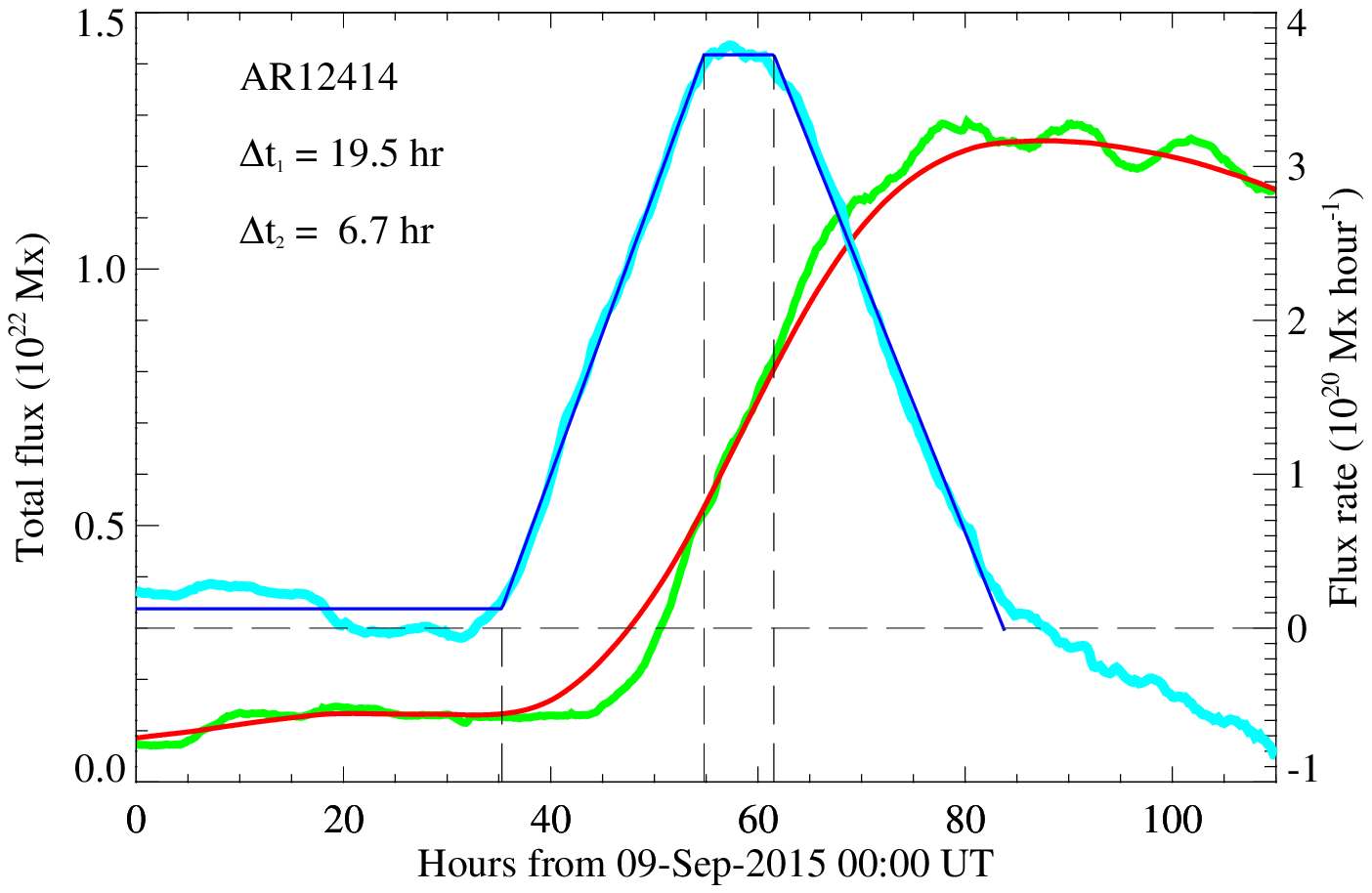}
  	}
  	\vspace{-0.35\textwidth}   
  	\centerline{\Large \bf     
  		\hspace{0.0 \textwidth} \color{white}{(c)}
  		\hspace{0.415\textwidth}  \color{white}{(d)}
  		\hfill}
  	\vspace{0.31\textwidth}    
  	\centerline{\hspace*{0.015\textwidth}
  		\includegraphics[width=0.515\textwidth,clip=]{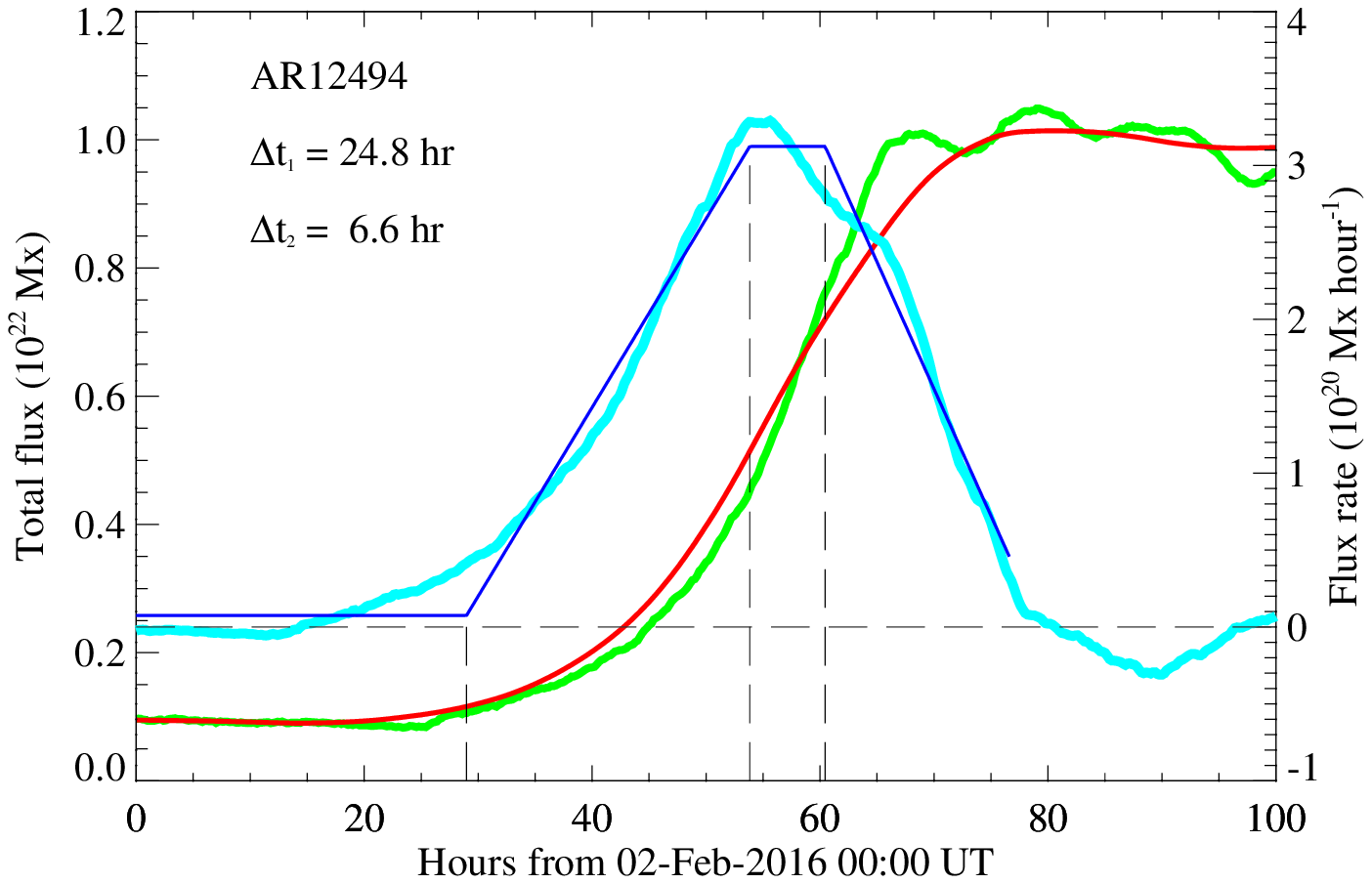}
  		\hspace*{-0.03\textwidth}
  		\includegraphics[width=0.515\textwidth,clip=]{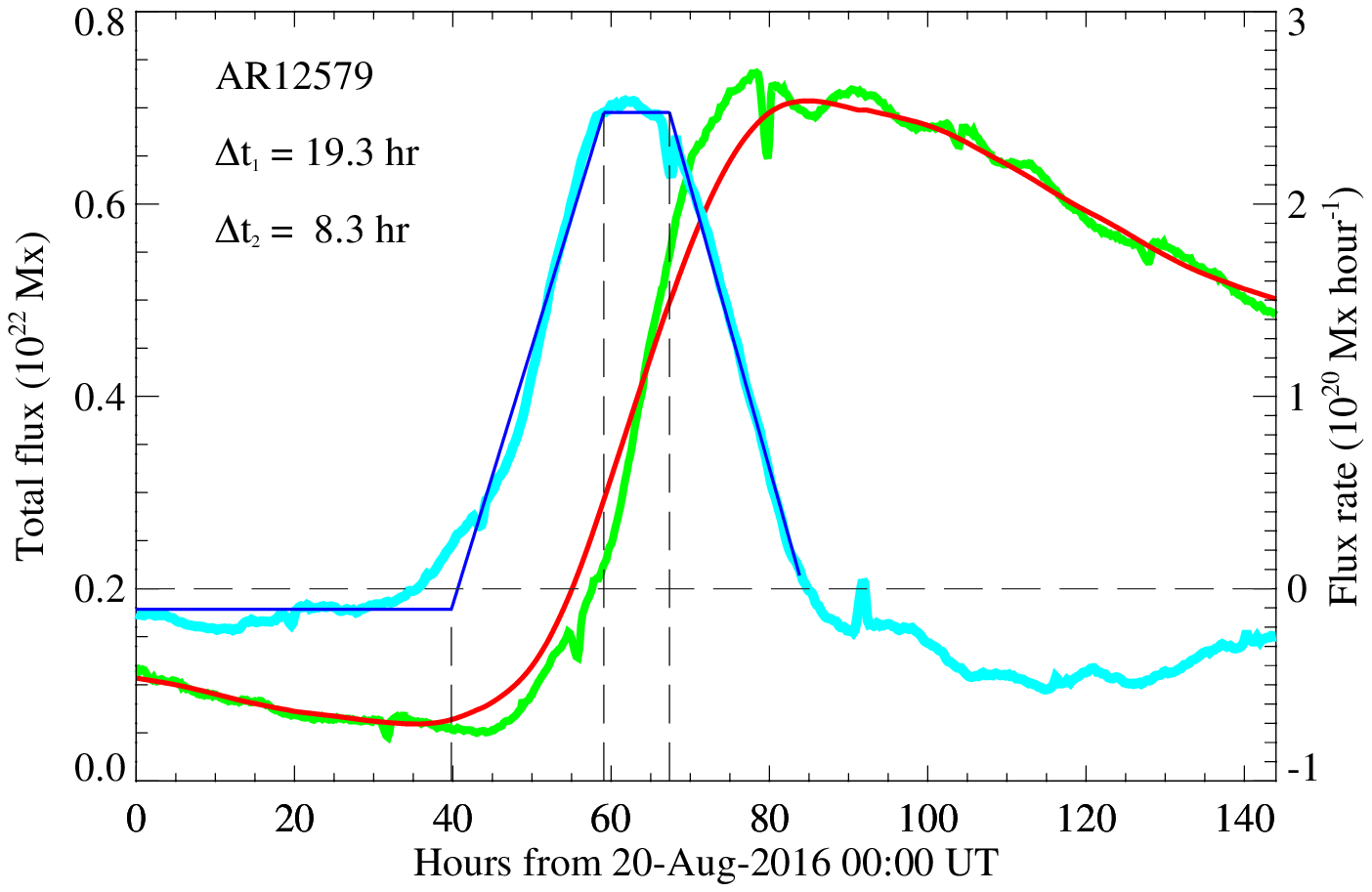}
  	}
  	\vspace{-0.35\textwidth}   
  	\centerline{\Large \bf     
  		\hspace{0.0 \textwidth} \color{white}{(c)}
  		\hspace{0.415\textwidth}  \color{white}{(d)}
  		\hfill}
  	\vspace{0.31\textwidth}    
  	
  	\caption{Time profiles of the total flux before (green curve) and after (red curve) the 24-hour filtering, the flux growth rate (turquoise curves), and the piecewise continuous linear fit (blue line segments) to the flux growth rate profiles; the three vertical dashed lines mark the start of the $\Delta t\textsubscript{1}$ interval, the start of the $\Delta t\textsubscript{2}$ interval, and the end of the $\Delta t\textsubscript{2}$ interval. The duration of $\Delta t\textsubscript{1}$ and $\Delta t\textsubscript{2}$ in hours are noted in the upper left corner of each plot. The horizontal dashed line shows the zero-level of the flux growth rate function, $R(t)$. Here we show  typical examples of emerging regions with a short (\textless13 hours) $\Delta t\textsubscript{2}$ interval and a high magnitude of the normalized flux growth rate ($R\textsubscript{N}$\textgreater0.024 hour\textsuperscript{-1}).
  	}
  	\label{Fig1}
  \end{figure}

\begin{figure}    
  	\centerline{\hspace*{0.015\textwidth}
  		\includegraphics[width=0.515\textwidth,clip=]{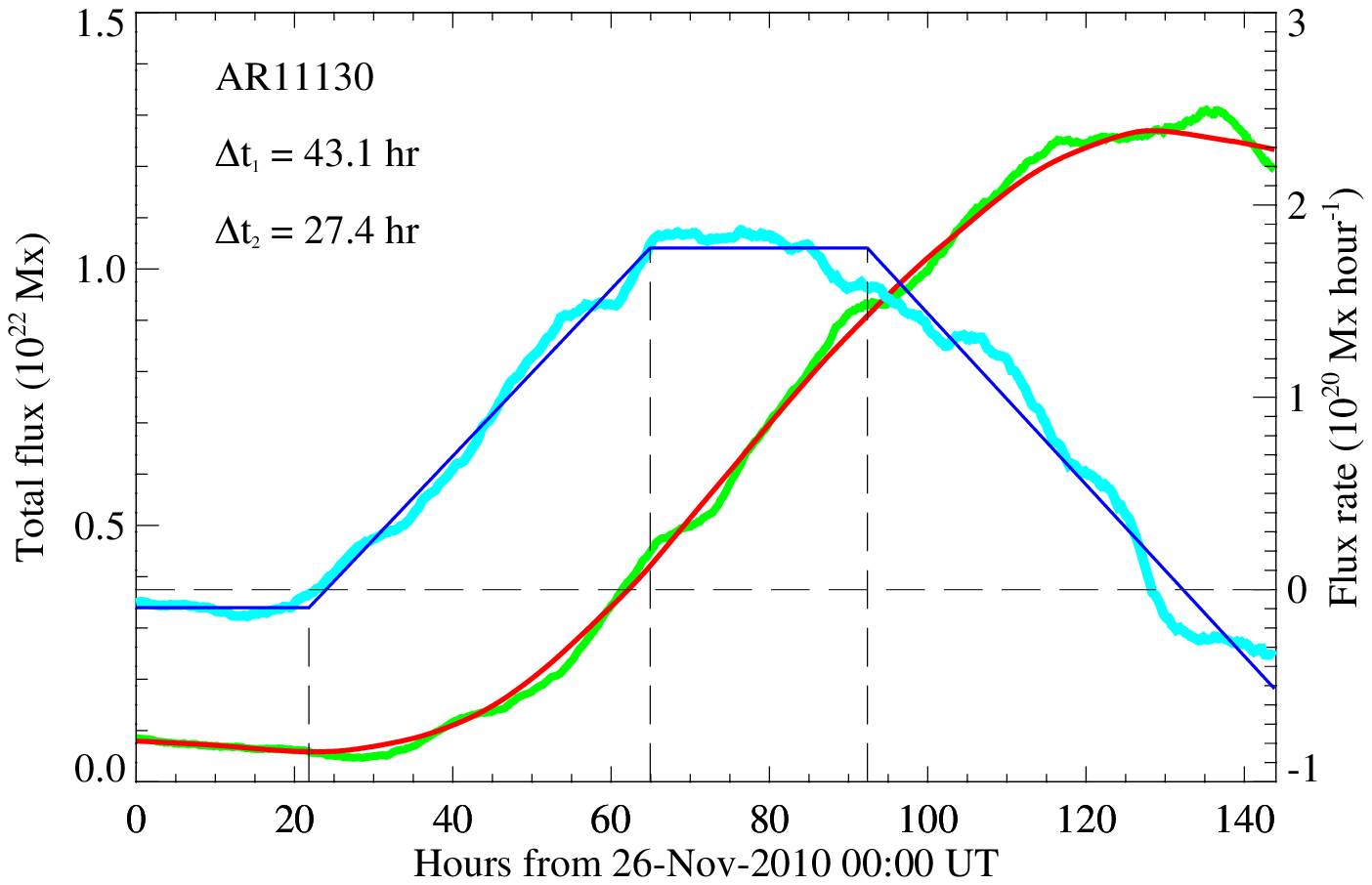}
  		\hspace*{-0.03\textwidth}
  		\includegraphics[width=0.515\textwidth,clip=]{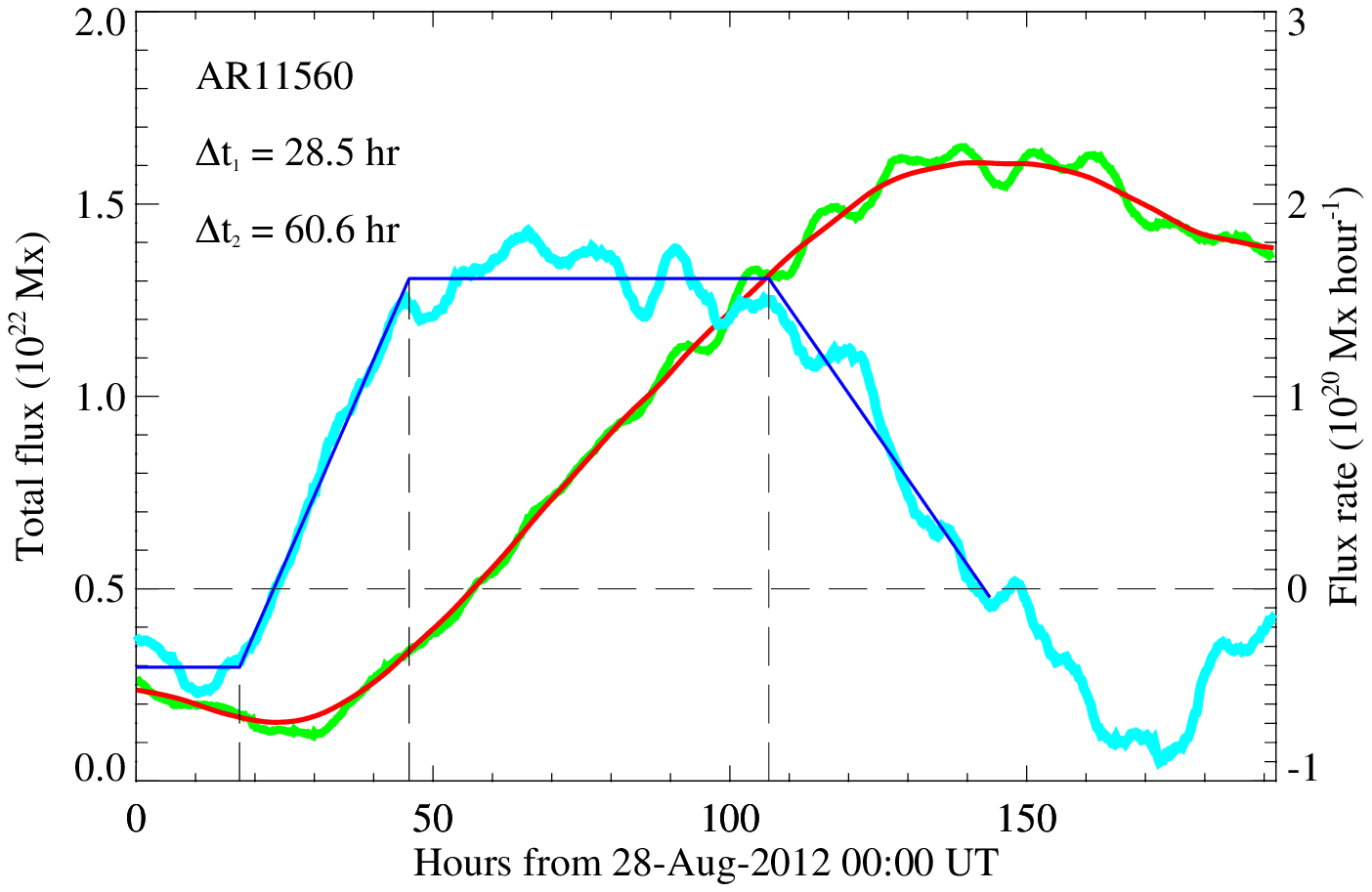}
  	}
  	\vspace{-0.35\textwidth}   
  	\centerline{\Large \bf     
  		\hspace{0.0 \textwidth}  \color{white}{(a)}
  		\hspace{0.415\textwidth}  \color{white}{(b)}
  		\hfill}
  	\vspace{0.31\textwidth}    
  	\centerline{\hspace*{0.015\textwidth}
  		\includegraphics[width=0.515\textwidth,clip=]{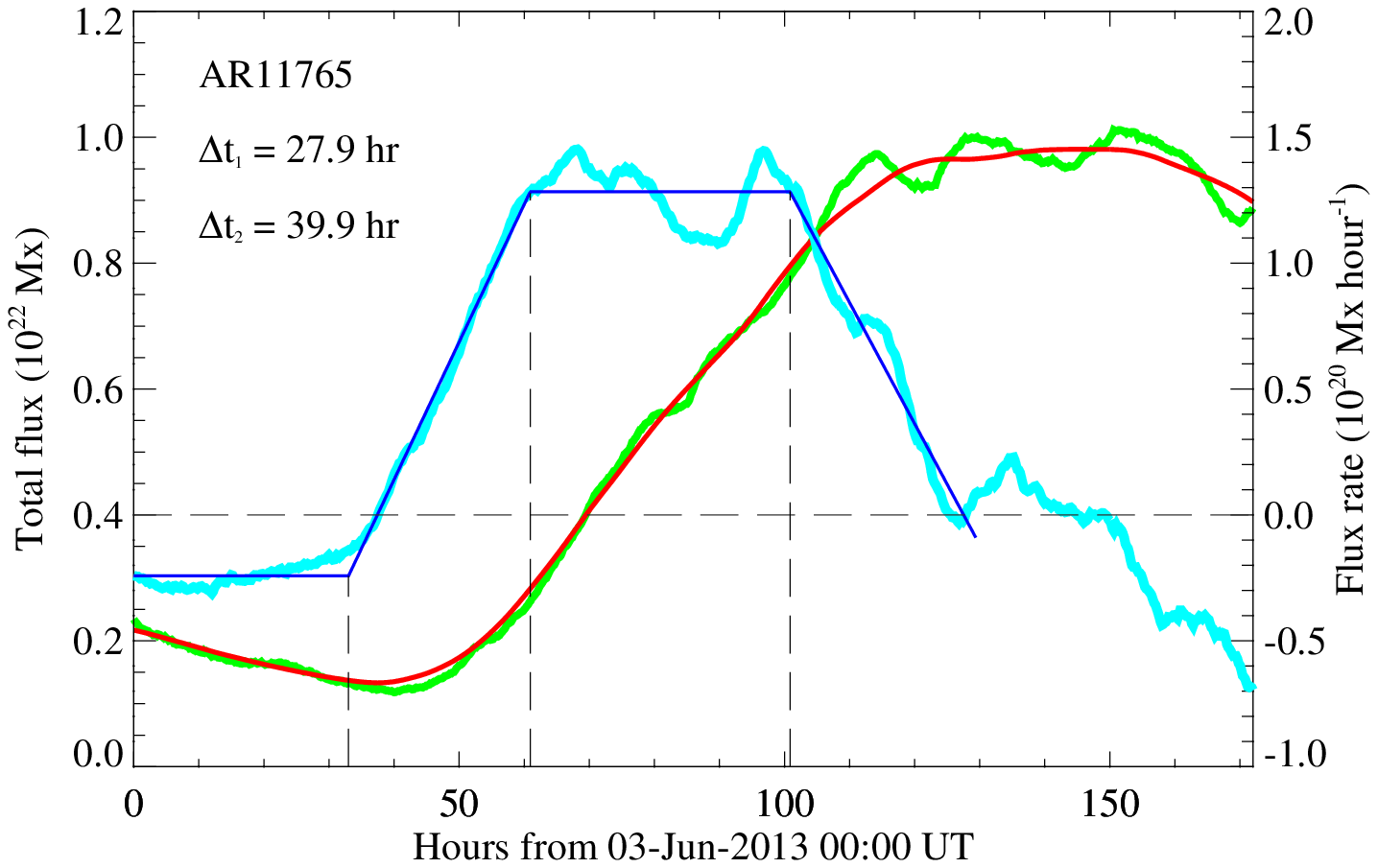}
  		\hspace*{-0.03\textwidth}
  		\includegraphics[width=0.515\textwidth,clip=]{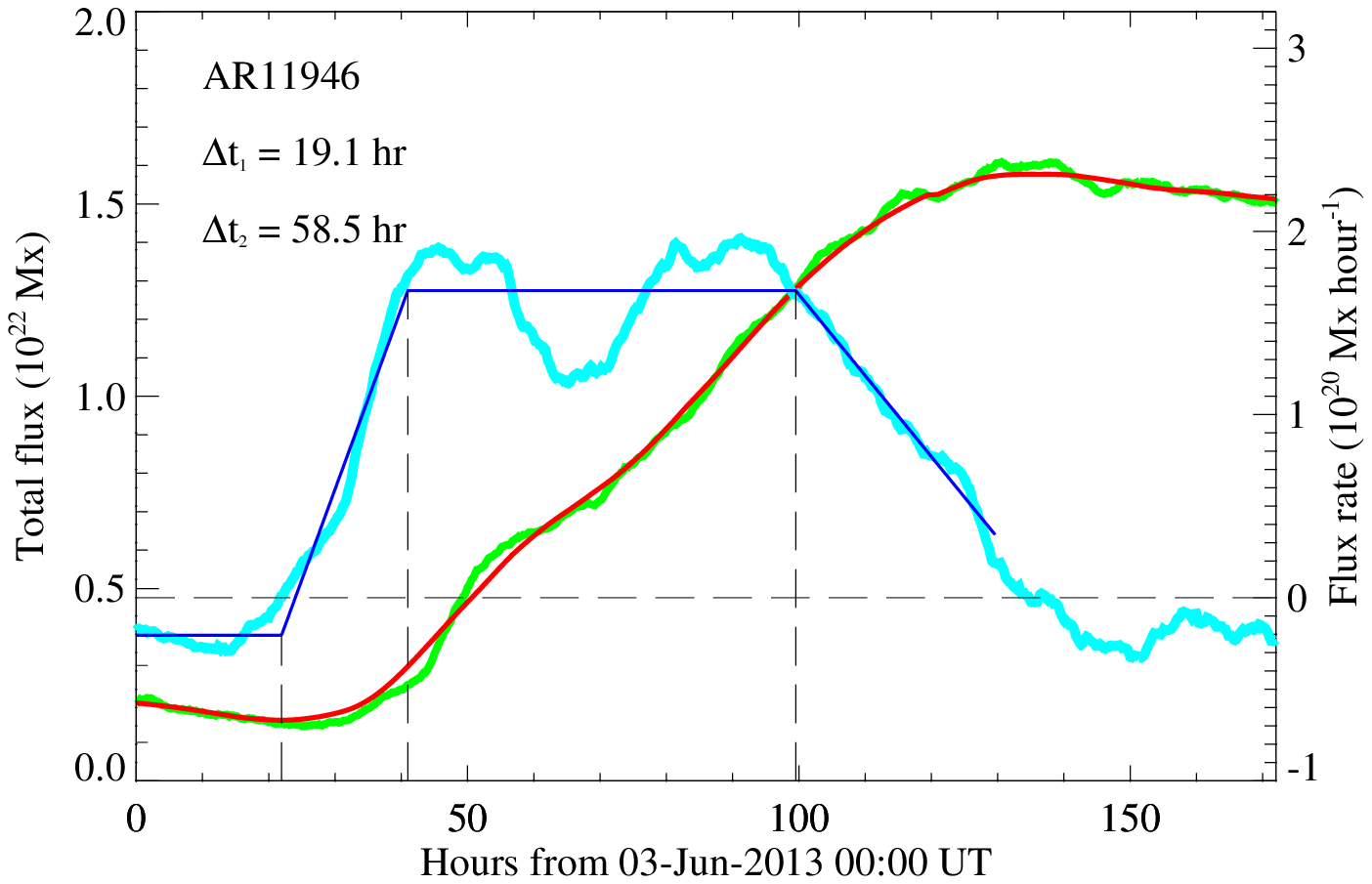}
  	}
  	\vspace{-0.35\textwidth}   
  	\centerline{\Large \bf     
  		\hspace{0.0 \textwidth} \color{white}{(c)}
  		\hspace{0.415\textwidth}  \color{white}{(d)}
  		\hfill}
  	\vspace{0.31\textwidth}    
  	\centerline{\hspace*{0.015\textwidth}
  		\includegraphics[width=0.515\textwidth,clip=]{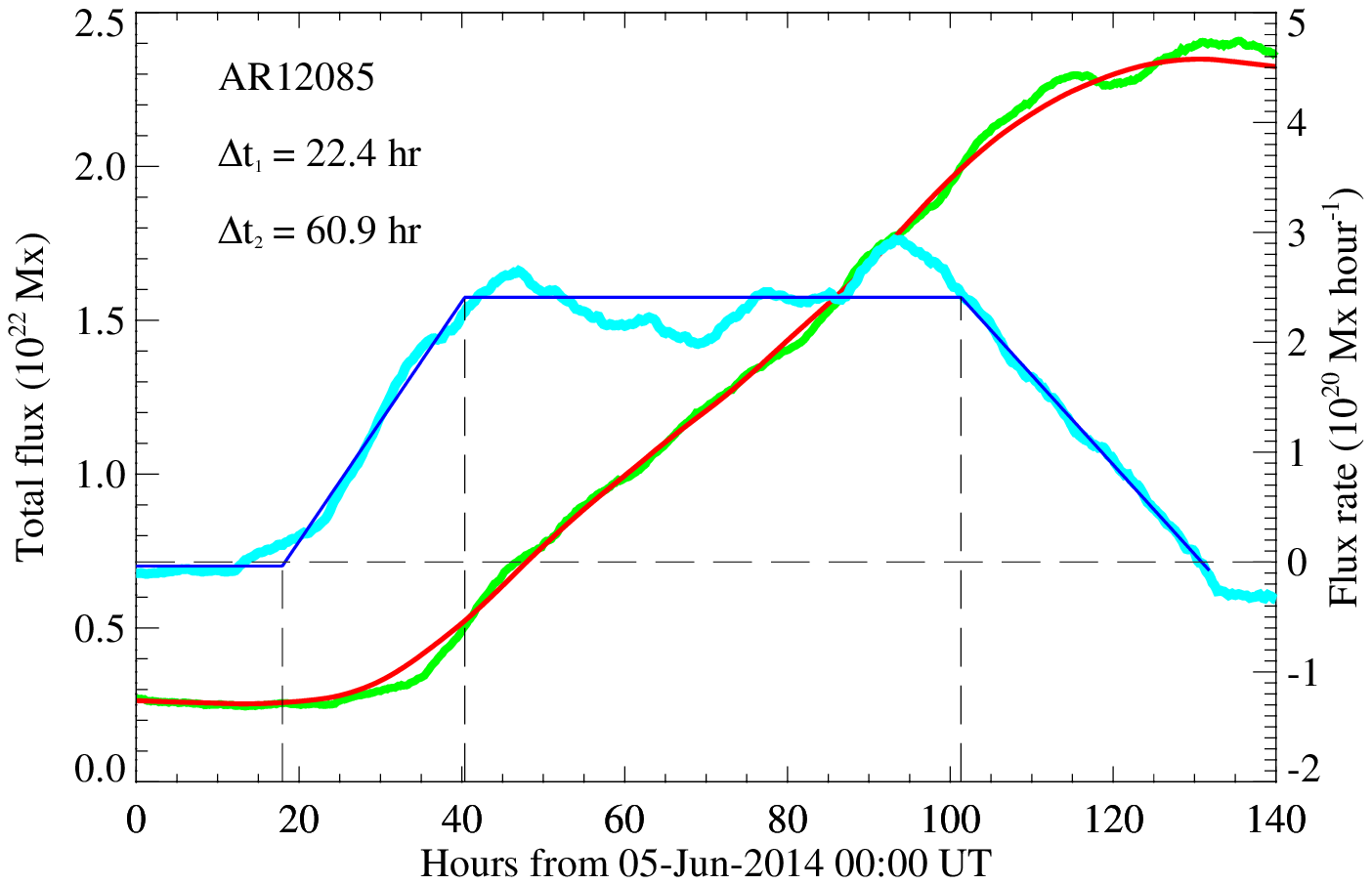}
  		\hspace*{-0.03\textwidth}
  		\includegraphics[width=0.515\textwidth,clip=]{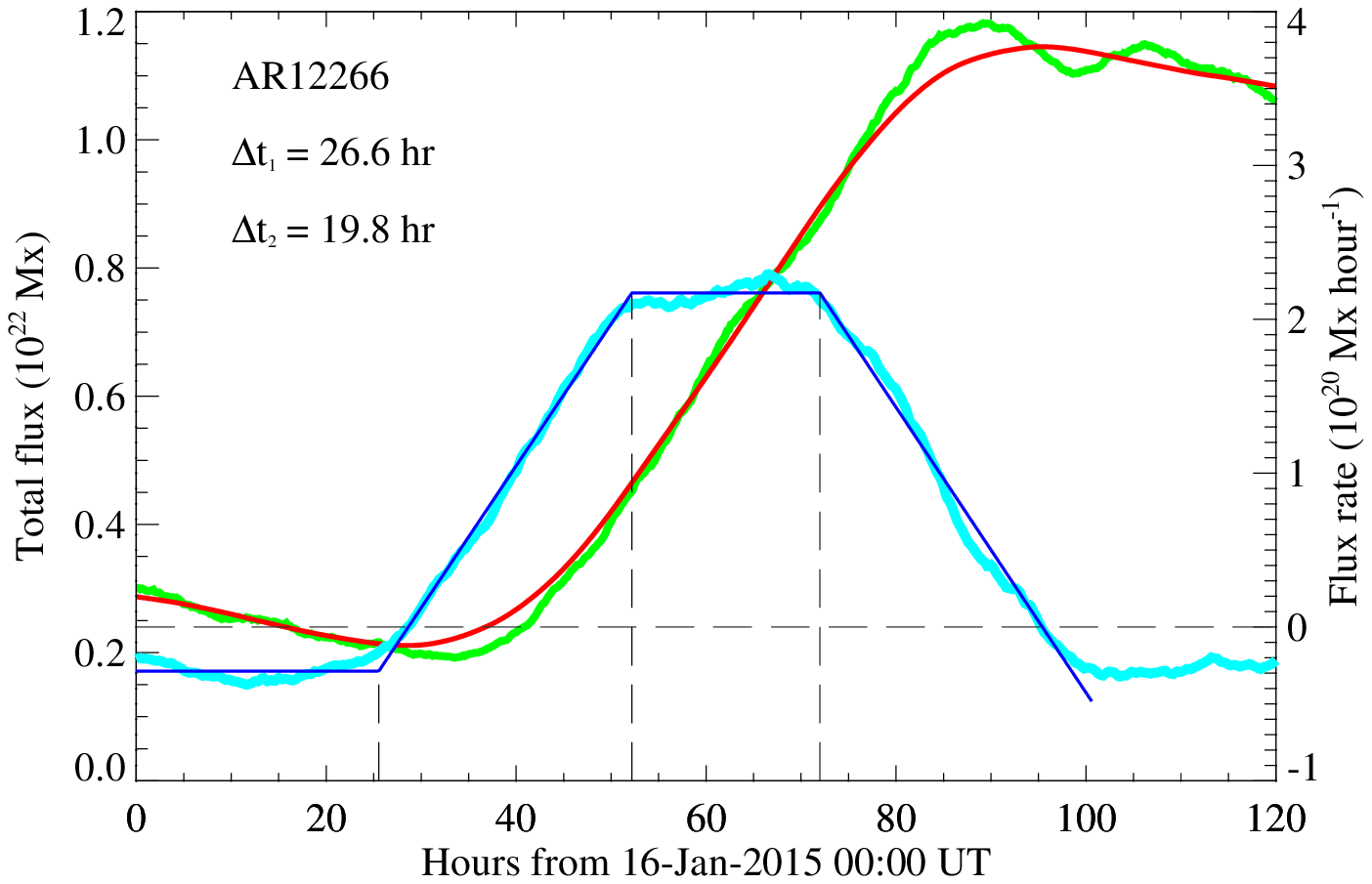}
  	}
  	\vspace{-0.35\textwidth}   
  	\centerline{\Large \bf     
  		\hspace{0.0 \textwidth} \color{white}{(c)}
  		\hspace{0.415\textwidth}  \color{white}{(d)}
  		\hfill}
  	\vspace{0.31\textwidth}    
  	
  	\caption{The same as in Figure \ref{Fig1} for typical examples of  emerging regions with a long  (\textgreater13 hours) $\Delta t\textsubscript{2}$ interval and a low  magnitude of the normalized flux growth rate ($R\textsubscript{N}$\textless0.024 hour\textsuperscript{-1}).
  	}
  	\label{Fig2}
  \end{figure}

\begin{figure}    
  	\centerline{\hspace*{0.015\textwidth}
  		\includegraphics[width=0.515\textwidth,clip=]{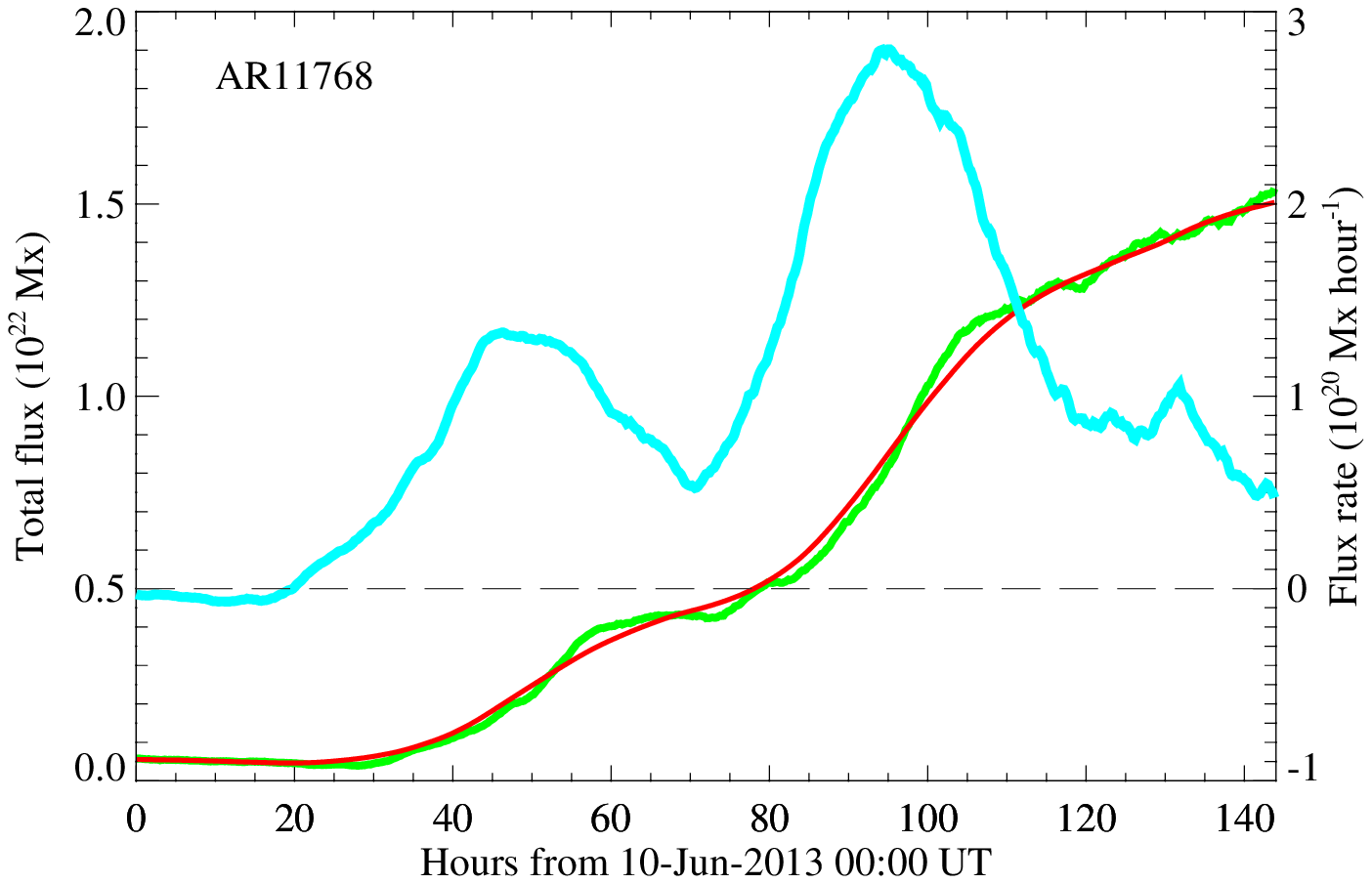}
  		\hspace*{-0.03\textwidth}
  		\includegraphics[width=0.515\textwidth,clip=]{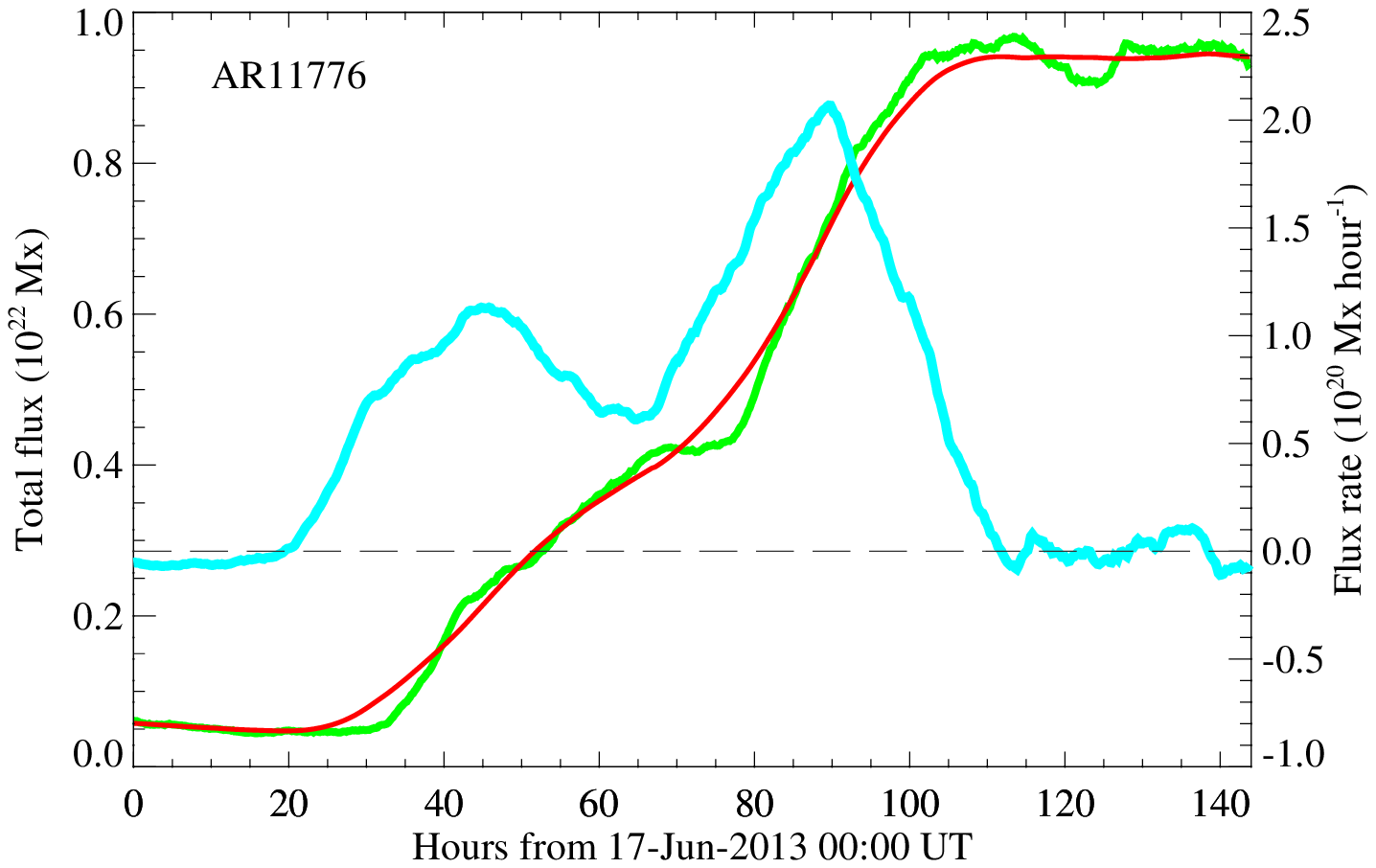}
  	}
  	\vspace{-0.35\textwidth}   
  	\centerline{\Large \bf     
  		\hspace{0.0 \textwidth}  \color{white}{(a)}
  		\hspace{0.415\textwidth}  \color{white}{(b)}
  		\hfill}
  	\vspace{0.31\textwidth}    
  	\centerline{\hspace*{0.015\textwidth}
  		\includegraphics[width=0.515\textwidth,clip=]{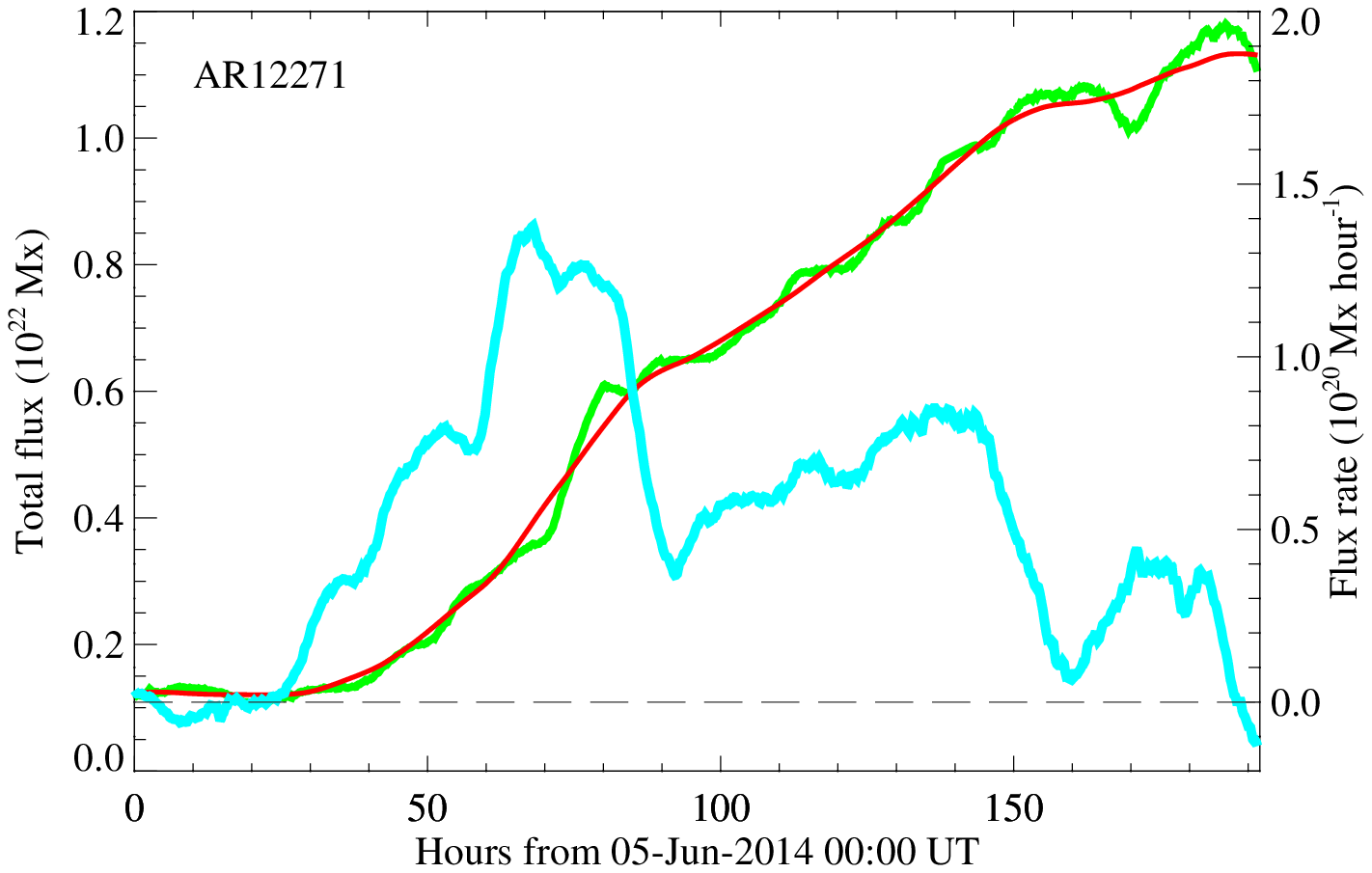}
  		\hspace*{-0.03\textwidth}
  		\includegraphics[width=0.515\textwidth,clip=]{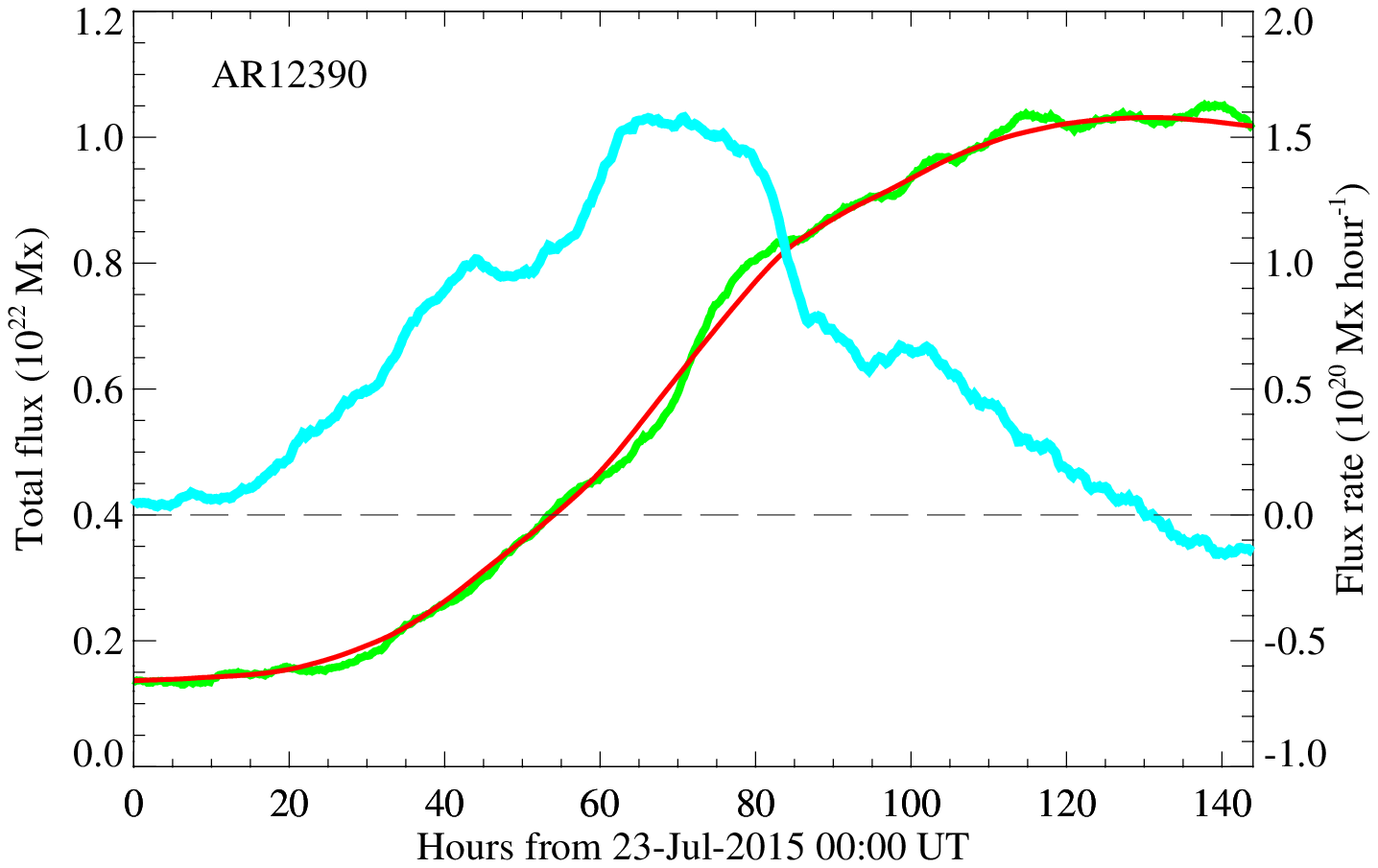}
  	}
  	\vspace{-0.35\textwidth}   
  	\centerline{\Large \bf     
  		\hspace{0.0 \textwidth} \color{white}{(c)}
  		\hspace{0.415\textwidth}  \color{white}{(d)}
  		\hfill}
  	\vspace{0.31\textwidth}    
  	
  	\caption{The same as in Figure \ref{Fig1} for four ARs, where the piecewise continuous linear fitting algorithm failed.
  	}
  	\label{Fig3}
  \end{figure}

For this study, we used line-of-sight magnetic field data taken by the \textit{Helioseismic and Magnetic Imager} (HMI, \cite{Scherrer2012, Schou2012}) on board the \textit{Solar Dynamics Observatory} (SDO, \cite{Pesnell2012}). Continuous 5-7 day long data series for 42 emerging ARs were selected for the study. We chose only isolated ARs, emerging amidst the quiet Sun. The list of the ARs, the analyzed  time interval, the longitude of emergence, and calculated emergence parameters are listed in Table \ref{Table1}. We selected only those ARs that emerged at heliographic longitudes less than 60 degree (except a one event) to avoid errors that can be introduced by the foreshortening effect. All but one AR (NOAA AR 11678) first appeared in the eastern hemisphere which allowed us to track the emergence process during an extended period of time. The 720-second magnetograms were carefully aligned using the IDL Fast Fourier Transform algorithm to remove the solar rotation effect. The total unsigned flux was calculated from each magnetogram as a sum of absolute flux values measured at each pixel. Only those pixels with the modulus of the flux density twice exceeding the noise level were taken into account. The standard deviation of noise for HMI magnetograms is about 5-6 G \citep{Liu2012}. The observed total unsigned flux was then corrected for the foreshortening  effect by dividing it with the cosine of the heliographic longitude (see, \cite{Hagenaar2001}). The resulting time profiles of the total unsigned flux for selected ARs are shown in Figures \ref{Fig1}, \ref{Fig2}, and \ref{Fig3} with green curves. We observed an active region until the moment when the total unsigned flux saturates, \textit{i.e.}, its derivative changes the sign from positive to negative, see Figures \ref{Fig1}, \ref{Fig2}, \ref{Fig3}. We calculated the maximal total unsigned flux, $F\textsubscript{MAX}$, at the moment when the derivative equals zero.


\begin{table}
	\caption{Active regions under the study and the calculated emergence parameters}
	\label{Table1}
	\begin{tabular}{cccccccc}     
		\hline                   
		AR & Observation & Emerg. & $F\textsubscript{MAX}$  & $\Delta t\textsubscript{1}$ & $\Delta t\textsubscript{2}$ & $R\textsubscript{MAX}$ & $R\textsubscript{N}$\\
		no.	&  window  & long. 	 & 10\textsuperscript{22} Mx & hours  & hours & 10\textsuperscript{20} Mx hour\textsuperscript{-1}& hour\textsuperscript{-1}\\
		
		\hline
		
		11130   &   2010.11.26-12.01   &   E10   &    1.27   &   43.1  &  27.4  &  1.78(0.21)\tabnote{standard deviation values are shown in parenthesis} &  0.014 \\
		11158   &   2011.02.10-02.18   &   E22   &    2.49   &   20.8  &  02.2  &  4.97(0.01) &  0.020 \\
		11184   &   2011.04.01-04.05   &   E16   &    1.07   &   27.6  &  11.3  &  2.42(0.12) &  0.023 \\
		11416   &   2012.02.07-02.14   &   E48   &    1.86   &   50.6  &  08.2  &  3.30(0.06) &  0.018 \\
		11422   &   2012.02.16-02.22   &   E22   &    1.39   &   42.3  &  00.1  &  3.60(0.07) &  0.026 \\
		11455   &   2012.04.10-04.15   &   E31   &    0.80   &   44.8  &  21.8  &  1.16(0.16) &  0.015 \\
		11554   &   2012.08.22-08.28   &   E43   &    0.80   &   31.0  &  05.0  &  2.12(0.08) &  0.027 \\
		11560   &   2012.08.28-09.04   &   E51   &    1.61   &   28.5  &  60.6  &  1.63(0.18) &  0.010 \\
		11620   &   2012.11.23-11.28   &   E01   &    2.13   &   21.6  &  09.4  &  3.58(0.20) &  0.017 \\
		11660   &   2013.01.18-01.24   &   E14   &    1.46   &   24.4  &  13.1  &  2.04(0.18) &  0.014 \\
		11670   &   2013.02.06-02.12   &   E50   &    1.01   &   28.7  &  31.8  &  1.14(0.06) &  0.011 \\
		11675   &   2013.02.15-02.20   &   E57   &    0.58   &   23.0  &  04.3  &  1.70(0.09) &  0.029 \\
		11678   &   2013.02.15-02.20   &   W23   &    1.90   &   26.3  &  02.0  &  4.38(0.07) &  0.023 \\
		11696   &   2013.03.10-03.16   &   E48   &    0.98   &   15.4  &  38.9  &  1.37(0.21) &  0.014 \\
		11726   &   2013.04.17-04.24   &   E25   &    2.62   &   38.6  &  18.3  &  4.25(0.23) &  0.016 \\
		11764   &   2013.06.01-06.05   &   E04   &    0.61   &   22.9  &  08.4  &  1.86(0.03) &  0.030 \\
		11765   &   2013.06.03-06.11   &   E39   &    0.98   &   27.9  &  39.9  &  1.29(0.17) &  0.013 \\
		11781   &   2013.06.27-07.01   &   E21   &    0.84   &   39.7  &  37.7  &  1.00(0.16) &  0.012 \\
		11837   &   2013.08.30-09.05   &   E02   &    1.04   &   33.5  &  35.1  &  1.23(0.11) &  0.012 \\
		11855   &   2013.09.29-10.04   &   E38   &    0.90   &   33.7  &  12.9  &  1.83(0.02) &  0.020 \\
		11916   &   2013.12.03-12.09   &   E18   &    2.25   &   24.8  &  25.0  &  2.23(0.19) &  0.010 \\
		11928   &   2013.12.15-12.21   &   E35   &    2.12   &   27.1  &  04.8  &  3.57(0.05) &  0.017 \\
		11946   &   2014.01.03-01.10   &   E50   &    1.58   &   19.1  &  58.5  &  1.68(0.26) &  0.011 \\
		12036   &   2014.04.11-04.19   &   E55   &    1.79   &   42.3  &  03.6  &  3.21(0.10) &  0.018 \\
		12085   &   2014.06.05-06.11   &   E41   &    2.35   &   22.4  &  60.9  &  2.41(0.25) &  0.010 \\
		12175   &   2014.09.22-09.28   &   E16   &    2.78   &   23.8  &  26.7  &  3.38(0.35) &  0.012 \\
		12203   &   2014.10.31-11.06   &   E22   &    1.06   &   21.5  &  06.6  &  2.86(0.14) &  0.027 \\
		12257   &   2015.01.03-01.11   &   E28   &    2.00   &   39.5  &  33.3  &  2.63(0.36) &  0.013 \\
		12266   &   2015.01.16-01.22   &   E37   &    1.15   &   26.6  &  19.8  &  2.17(0.06) &  0.019 \\
		12273   &   2015.01.23-01.30   &   E22   &    0.51   &   20.2  &  12.6  &  1.49(0.06) &  0.029 \\
		12275   &   2015.01.23-01.30   &   E00   &    0.77   &   25.6  &  03.6  &  2.34(0.15) &  0.030 \\
		12353   &   2015.05.20-05.26   &   E24   &    0.51   &   11.9  &  64.5  &  0.56(0.07) &  0.011 \\
		12414   &   2015.09.09-09.14   &   E01   &    1.25   &   19.5  &  06.7  &  3.80(0.16) &  0.030 \\
		12494   &   2016.02.02-02.09   &   E36   &    1.14   &   25.8  &  06.6  &  3.12(0.13) &  0.027 \\
		12579   &   2016.08.20-08.27   &   E37   &    0.71   &   19.3  &  08.3  &  2.55(0.12) &  0.036 \\
		12597   &   2016.09.20-09.27   &   E29   &    0.63   &   19.6  &  28.3  &  1.03(0.13) &  0.016 \\
		\\
		11682   &   2013.02.24-03.02   &   E34   &    1.24   &    -    &   -    &      -      &    -   \\
		11768   &   2013.06.10-06.15   &   E14   &    1.50   &    -    &   -    &      -      &    -   \\
		11776   &   2013.06.17-06.22   &   E24   &    0.95   &    -    &   -    &      -      &    -   \\
		12003   &   2014.03.08-03.13   &   E18   &    1.29   &    -    &   -    &      -      &    -   \\
		12271   &   2015.01.23-01.30   &   E43   &    1.13   &    -    &   -    &      -      &    -   \\
		12390   &   2015.07.23-07.30   &   E68   &    1.03   &    -    &   -    &      -      &    -   \\

		\hline
	\end{tabular}
\end{table}

SDO/HMI magnetic data exhibit artificial oscillations with а period of 24 hours caused by instrumental effects \citep{Liu2012, Smirnova2013, Kutsenko2016}. Following the routine suggested by \cite{Kutsenko2016}, we performed  low-frequency filtering to remove these artificial oscillations and the resulting profiles, $F(t)$, are shown in Figures \ref{Fig1}, \ref{Fig2}, and \ref{Fig3} with red curves. One can see that artificial flux oscillations are not negligible. These projection-corrected and filtered time profiles of the total unsigned flux, $F(t)$, will be used in our study to calculate the flux growth rate. For sake of simplicity, further in the text we will refer to it as the total flux, $F(t)$.

The flux growth rate was calculated as a time derivative, $R(t)$, of the total  flux, $F(t)$: $R(t)=d(F(t))/dt$. Since the time derivative does not depend on any constant component of the flux itself, we ignored the contribution from the quiet Sun areas within a magnetogram, contrary to a routine suggested in \cite{vanDrielGesztelyi2003}. Time variations of the flux growth rate are presented in Figures \ref{Fig1}, \ref{Fig2}, and \ref{Fig3} with turquoise curves.

Having the time profiles of the $R(t)$ function during the emergence, for each AR we performed a piecewise continuous linear fitting under the following requirements: the fit should consist of four linear pieces; the first and third segments are horizontal, whereas the second and the fourth segments may have an arbitrary slope. The parameters of the linear pieces are calculated under the requirement of the minimum mean square deviations from the observed $R(t)$ curve. The start time was set to the first magnetogram acquisition time, and the end time of the fitting interval in the most of the cases was chosen as the moment where $R(t)=0$. In  cases of complex $R(t)$ profiles, the end time was modified in order to preserve the quality of the fit. The fitting procedure gave us two characteristic time intervals: $\Delta t\textsubscript{1}$ –- the interval of continuously increasing flux increment, and $\Delta t\textsubscript{2}$ –- the interval of quasi-constant flux increment (see Table \ref{Table1}, 5\textsuperscript{th} and 6\textsuperscript{th} columns, respectively). The latter is followed by an interval of decreasing flux increment. We were successful in fitting 36 out of 42 ARs, while the remaining 6 ARs display a complex $R(t)$ profile (Figure \ref{Fig3}), so that the fitting algorithm failed. We include these ARs in Table \ref{Table1}, however, no fitting parameters are provided in these cases.

As it follows from Figures \ref{Fig1} and \ref{Fig2}, during the $\Delta t\textsubscript{2}$-interval, the flux growth rate, $R(t)$, undulates around the flat fitline reaching its maximum at some moment. The undulations can be rather low (see, \textit{e.g.}, Figure \ref{Fig2}, AR 12266) and high (\textit{e.g.}, AR 11946 in Figure \ref{Fig2}), and the absolute maximum of $R(t)$ does not seem to adequately describe the AR emergence process. Instead, we accept the averaged over the $\Delta t\textsubscript{2}$ value of $R(t)$ as a characteristic of the maximum flux growth rate of an AR. Their magnitudes, calculated as $R\textsubscript{MAX}=<R(t)\vert$ \textsubscript{$\Delta t\textsubscript{2}$}$>$ are listed in the 7\textsuperscript{th} column of Table \ref{Table1}.

As can be seen from Table \ref{Table1}, the $R\textsubscript{MAX}$ varies within a broad range of values and, presumably, depends on the saturated (maximal) total unsigned flux, $F\textsubscript{MAX}$ (4\textsuperscript{th} column in Table \ref{Table1}), determined at the time when $R(t)=0$. To compare the flux growth rate in strong and weak active regions, we calculated a flux growth rate of an AR normalized by the maximal total flux of that AR: $R\textsubscript{N}=R\textsubscript{MAX}/F\textsubscript{MAX}$. This parameter listed in the 8\textsuperscript{th} column of Table \ref{Table1} indicates the fraction of the maximal total flux added to the AR per hour during the $\Delta t\textsubscript{2}$-stage.

\section{Data Analysis}
\label{sec-DataAn}

\begin{figure}     
  	\centerline{\includegraphics[width=1.0\textwidth,clip=]{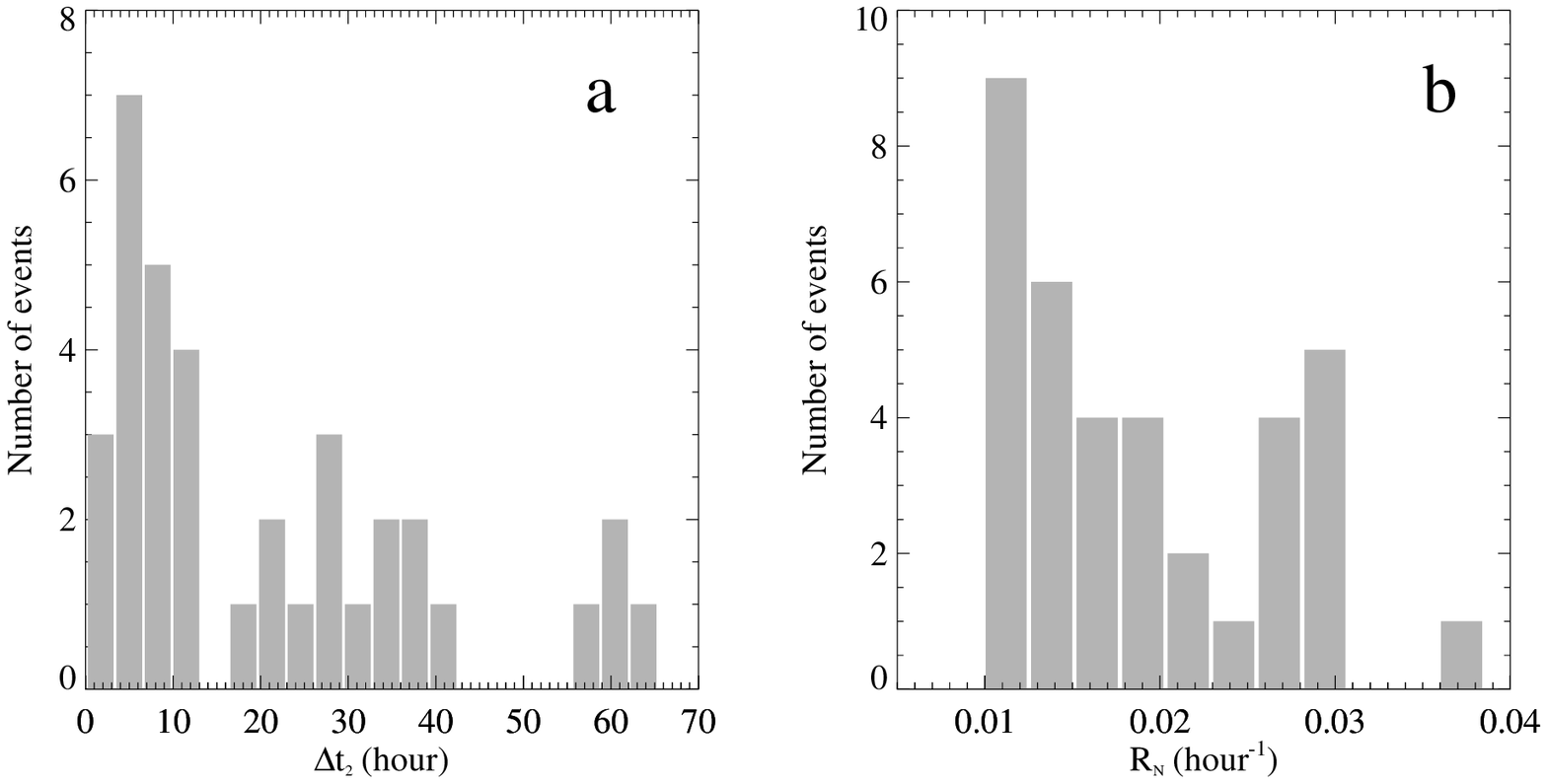}
  	}
  	\caption{
  		Distribution of $\Delta t\textsubscript{2}$ intervals (a) and of the normalized flux growth rates, $R\textsubscript{N}$, (b) derived for 36 ARs. 
  	}
  	\label{Fig4}
  \end{figure}

\begin{figure}     
  	\centerline{\includegraphics[width=1.0\textwidth,clip=]{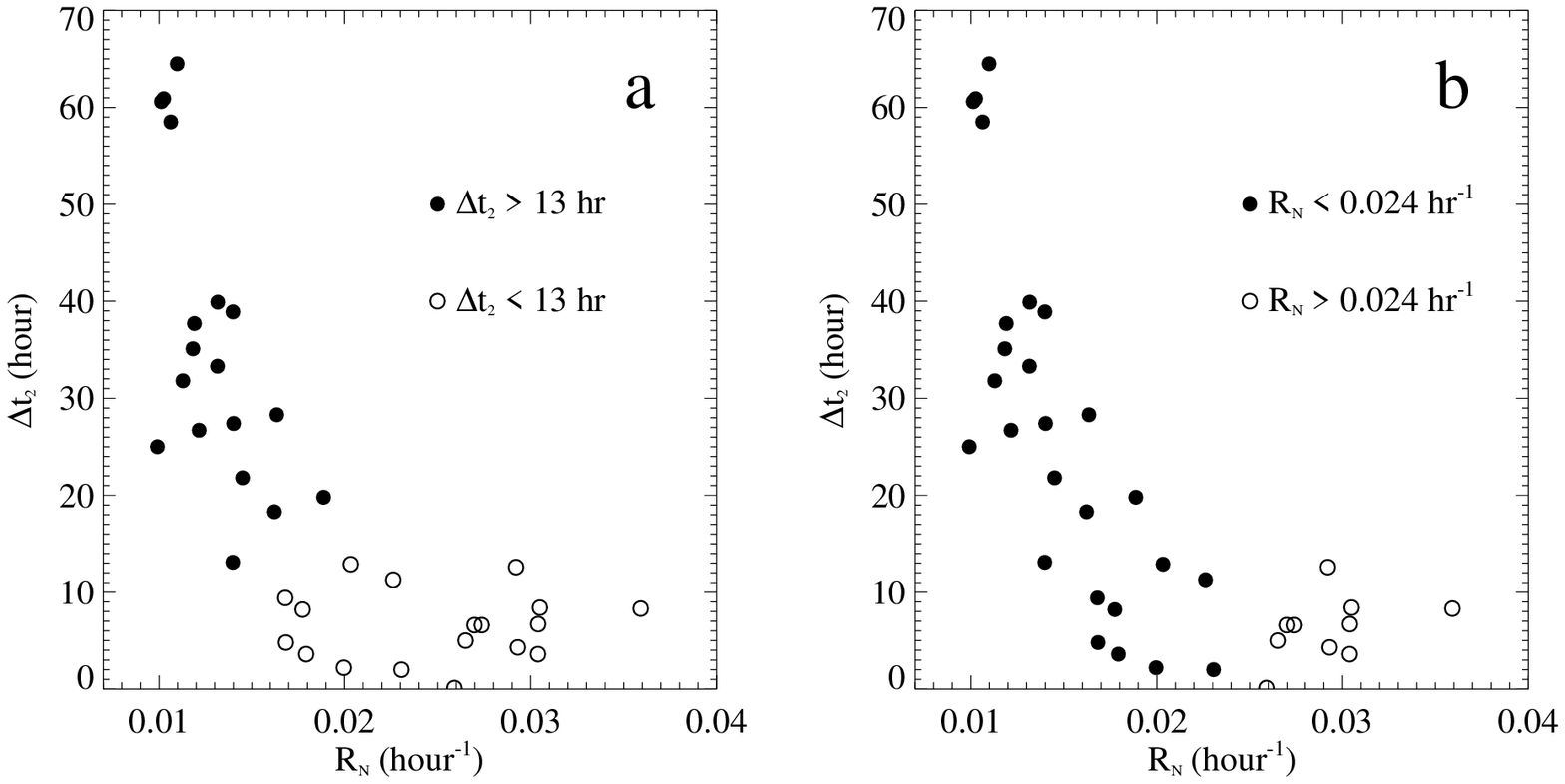}
  	}
  	\caption{
  		Plots of $\Delta t\textsubscript{2}$ \textit{versus} $R\textsubscript{N}$. a -– Events with  $\Delta t\textsubscript{2}$\textgreater13 hours ($\Delta t\textsubscript{2}$\textless13) hours  are marked with filled (open)  circles. b -– Events with $R\textsubscript{N}$\textless0.024 hour\textsuperscript{-1}  ($R\textsubscript{N}$\textgreater0.024 hour\textsuperscript{-1}) are marked with filled (open) circles. Note that in panel b the symbol coding has changed for 8 events.
  	}
  	\label{Fig5}
  \end{figure}

The $\Delta t\textsubscript{2}$ and $R\textsubscript{N}$ histograms using data from 36 ARs are shown in Figures \ref{Fig4}a and \ref{Fig4}b, respectively. Figure \ref{Fig4}a shows a gap at approximately 13 hours, while in Figure \ref{Fig4}b we see a similar separation of the data points into two subsets with $R\textsubscript{N}$\textless0.024 hour\textsuperscript{-1} and $R\textsubscript{N}$\textgreater0.024 hour\textsuperscript{-1}. Are these $\Delta t\textsubscript{2}$ and $R\textsubscript{N}$ subsets independent? To answer the question, we plot $\Delta t\textsubscript{2}$ versus $R\textsubscript{N}$ in two ways. In Figure \ref{Fig5}a open circles represent cases with $\Delta t\textsubscript{2}$\textless13 hour, whereas in Figure \ref{Fig5}b open circles represent  cases with $R\textsubscript{N}$\textgreater0.024 hour\textsuperscript{-1}.  We see that only for 8 cases the symbol coding has changed to the opposite, while the symbol coding for the rest 28 cases remains unchanged. This means that for majority of ARs the $\Delta t\textsubscript{2}$ interval and  the normalized flux emergence rate, $R\textsubscript{N}$ are inversely proportional. For sake of simplicity further in the text the ARs with short $\Delta t\textsubscript{2}$ and high $R\textsubscript{N}$ will be referred to as "rapid" emergence events, and, correspondingly, the ARs with long $\Delta t\textsubscript{2}$ and low $R\textsubscript{N}$ will be called "gradual" emergence events.

\begin{figure}     
  	\centerline{\includegraphics[width=1.0\textwidth,clip=]{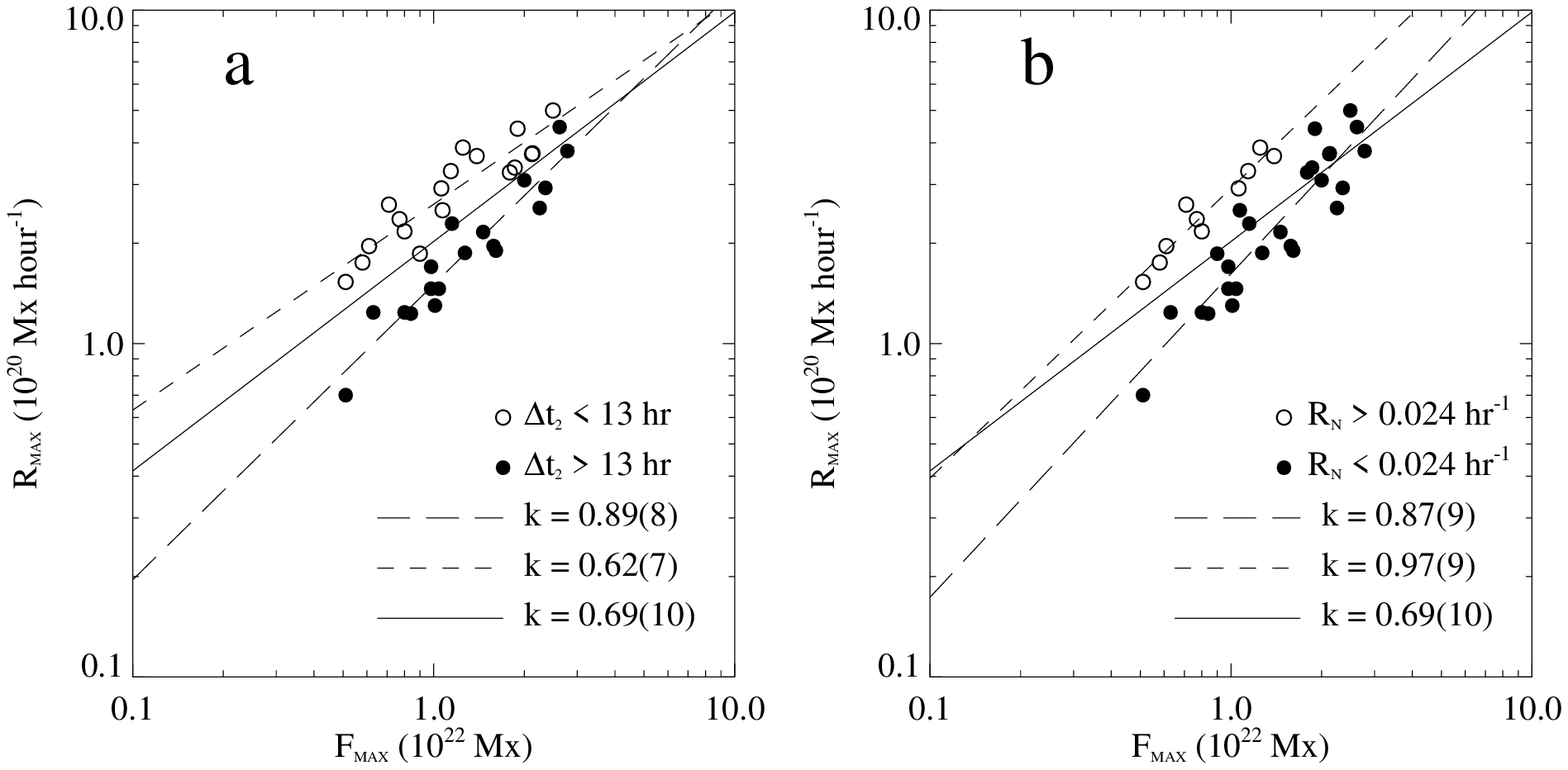}
  	}
  	\caption{
  		Flux growth rate $R\textsubscript{MAX}$ \textit{versus} the maximal total flux $F\textsubscript{MAX}$ for 36 ARs. Frames a and b differ in the segregation method. Symbol coding is the same as in Figure \ref{Fig5}. 
  	}
  	\label{Fig6}
  \end{figure}

There exists a well pronounced statistical dependence between the $F\textsubscript{MAX}$ and the flux growth rate, $R\textsubscript{MAX}$ (Figure \ref{Fig6}). When all 36 cases are considered, a linear regression in the double-logarithmic plot with the slope k=0.69$\pm$0.10 is well defined with Pearson correlation coefficient of 0.75.  When the "rapid" and "gradual" emergence cases are considered separately, again, the two subsets do not overlap for both methods of segregation, see panels a and b in Figure \ref{Fig6}. Moreover, the linear regressions show different slopes. Thus, the "gradually" emerging ARs (filled circles) display a steep slope $k$=0.89$\pm$0.08 for all $\Delta t\textsubscript{2}$\textgreater13 hours events and k=0.87$\pm$0.09 for all the $R\textsubscript{N}$\textless0.024 hour\textsuperscript{-1} events, and both differ significantly from the $k$=0.69 slope derived for the entire ensemble.  The "rapidly" emerging active regions (open circles) in both groups are tightly distributed around the linear (dashed line) fit (see Figure \ref{Fig6}), although the slope varies: $k$=0.62$\pm$0.07 for the $\Delta t\textsubscript{2}$\textless13 hours events and  $k$=0.97$\pm$0.09 for the $R\textsubscript{N}$\textgreater0.024 hour\textsuperscript{-1} events. Further studies with larger statistics are definitely needed to clarify the value of the slope.

\begin{figure}     
  	\centerline{\includegraphics[width=1.0\textwidth,clip=]{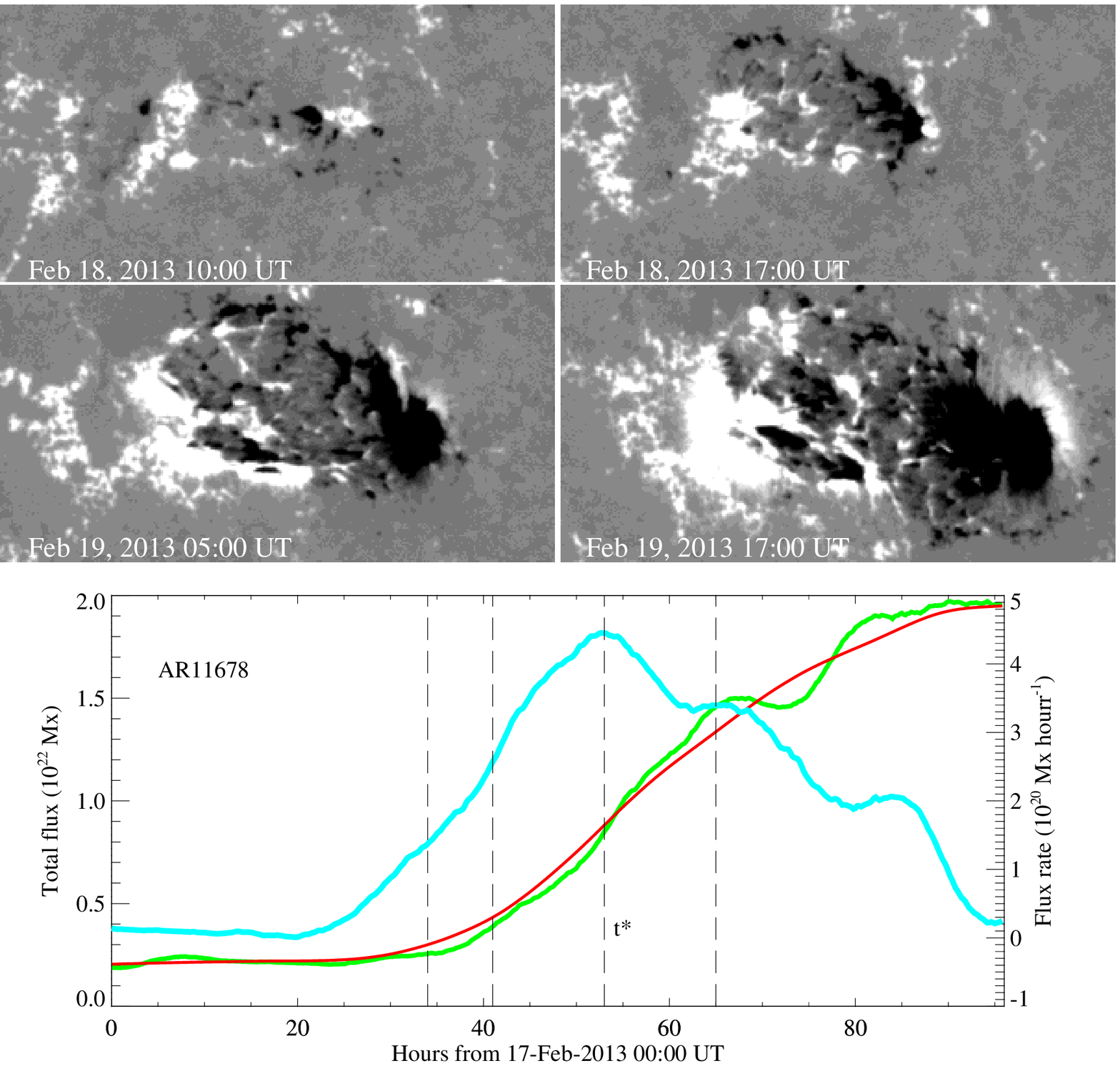}
  	}
  	\caption{
  		Top –- Four consecutive HMI magnetograms of NOAA AR 11678 scaled from 500 G (white) to -500 G (black). The size of  the field of view is 110 Mm $\times$ 55 Mm. North is to the top, west is to the right. Bottom -- Total unsigned flux before (green) and after (red) the 24-hour filtering and the flux growth rate $R(t)$ (turquoise). Vertical dashed lines denote the magnetogram acquisition times; t\textsuperscript{*} denotes the acquisition time of the third magnetogram near the maximum of $R(t)$. Animation of the emergence process is available in the supplementary materials. 
  	}
  	\label{Fig7}
  \end{figure}

An analysis of time series of magnetograms shows that for the "rapid" emergence ARs coherent appearance of a bundle of elongated mixed-polarity magnetic threads over a fast-expanding area is a characteristic feature (see Figure \ref{Fig7} and movies in the online version). At the very beginning of emergence, it is impossible to identify where the leading and following parts will be formed. An oval-like structure forms by the moment of maximal flux growth rate, $t\textsuperscript{*}$ (see bottom panel in Figure \ref{Fig7}).  After the peak time, $t\textsuperscript{*}$, the structure continues to emerge, however, in a different way: the leading and following parts are now well formed and separated from each other, and the mixed-polarity filaments/threads (presumably, "U" shaped parts of serpentine field lines) in the middle of the oval area become less ordered and either disappear at the location, or drift to the leading or following magnetic concentrations. The entire picture resembles the passage of a coherent flux tube through  the photosphere  in accordance with the "two-step" emergence model (see Conclusions and Discussion section).

\begin{figure}     
  	\centerline{\includegraphics[width=1.0\textwidth,clip=]{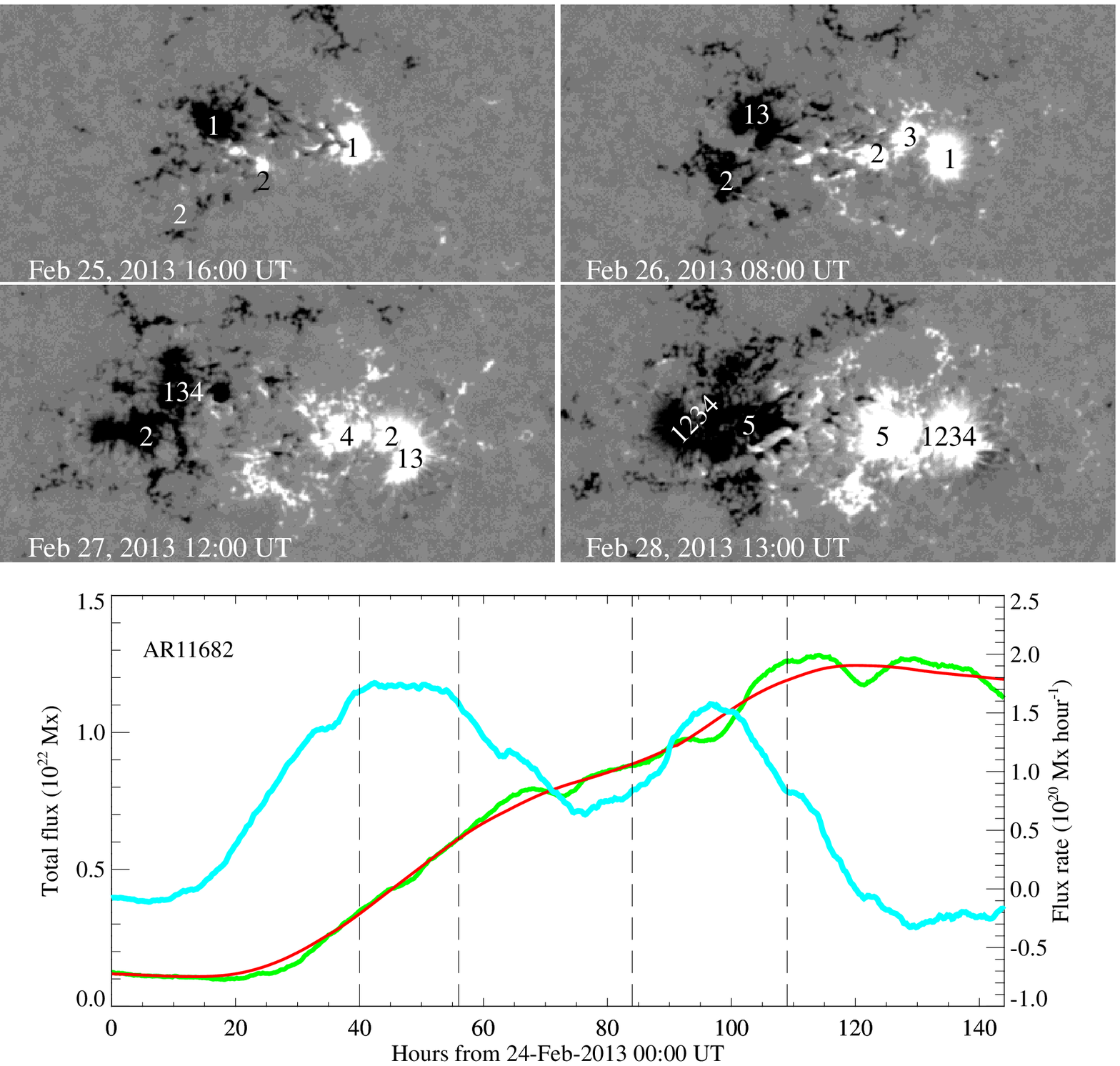}
  	}
  	\caption{
  		The same as in Figure \ref{Fig7} but for NOAA AR 11682. The size of field-of-view is 140 Mm $\times$ 70 Mm. The pairs of numbers (1 and 1), (2 and 2), \textit{etc.} denote several consecutively emerging magnetic dipoles. Notations are the same as in Figure \ref{Fig7}. Animation of the emergence process is available in the supplementary materials.
  	}
  	\label{Fig8}
  \end{figure}

Finally, in Figure \ref{Fig8} we show emergence of an AR where the piecewise fitting failed because of two well pronounced peaks in the $R(t)$-profile. The pairs of numbers (1 and 1), (2 and 2), \textit{etc.} denote five consecutively emerging magnetic dipoles which emerged one after another. These data  suggest that in case of "gradual" emergence, we may deal with a series of flux tubes, emerging along the same channel in the CZ, and appearing near the same place in the photosphere. Each one produces an $R(t)$-profile similar to that in Figure \ref{Fig7} (a single-peak profile), and their superposition in time results in the variety of $R(t)$-profiles shown in Figures \ref{Fig2} and \ref{Fig3}.

 \section{Conclusions and Discussion}
	  \label{sec-Conclusions}

In this paper, we  analyse the flux growth rate for 42 emerging ARs. The flux growth rate function, $R(t)$, was derived from the total unsigned flux, which was filtered to remove the artificial 24-hour oscillations in HMI data. This allowed us to obtain a more confident  estimate of the $R(t)$ than those reported in previous studies (\textit{e.g.}, \cite{Toriumi2014}).

A continuous piecewise linear fit to the $R(t)$-profiles allowed us to determine an interval $\Delta t\textsubscript{1}$ of increasing $R(t)$ and an interval $\Delta t\textsubscript{2}$ of nearly-constant $R(t)$ followed by the decrease of $R(t)$. This fitting routine failed to perform well for 6 ARs.

During the interval $\Delta t\textsubscript{2}$, the $R(t)$ function typically exhibited one or several local maxima. This allowed us to consider the averaged over the $\Delta t\textsubscript{2}$ growth rate as an estimate of the maximal value of the flux growth rate, $R\textsubscript{MAX}$. The 36 successfully fitted events form two subsets (with an 8 event overlap): i) ARs with short (\textless13 hours) $\Delta t\textsubscript{2}$ interval and a high (\textgreater0.024 hour\textsuperscript{-1}) normalized flux emergence rate, $R\textsubscript{N}$ which we call "rapid"-emergence events and ii) "gradual"-emergence events characterized by long (\textgreater13 hours) $\Delta t\textsubscript{2}$ interval and a low $R\textsubscript{N}$ (\textless0.024 hour\textsuperscript{-1}).

One more evidence for the separation is offered in the diagrams  $R\textsubscript{MAX}$ \textit{versus} $F\textsubscript{MAX}$, where the events from different subsets are not overlapped and each subset displays an individual power law. The power law index for the "rapid"-emergence events varies in a range of 0.62–-0.97, and in the range of 0.87-–0.89 for the "gradual"-emergence events. The power law index as derived for the entire ensemble of 36 events equals 0.69$\pm$0.10.

We found that the maximal flux growth rate $R\textsubscript{MAX}$ varies in a range of (0.5-5)$\times$10\textsuperscript{20} Mx hour\textsuperscript{-1} for active regions with the maximal total unsigned flux of (0.5-3)$\times$10\textsuperscript{22} Mx. Our data points (see Figure \ref{Fig6}) can be considered as an extension (toward larger fluxes and higher rates) of the diagram in Figure 5a in \cite{Otsuji2011}, where the flux growth rate varies in the range of (10\textsuperscript{18}-10\textsuperscript{20}) Mx hour\textsuperscript{-1} for regions with the total flux of 10\textsuperscript{18}-0.7$\times$10\textsuperscript{22} Mx. However, the power index $\alpha$ in the relation $R\textsubscript{MAX}\propto F\textsubscript{MAX}$\textsuperscript{$\alpha$} is higher (0.69) in our study as compared to that (0.57) reported in \cite{Otsuji2011}. The reason might be either a difference in the calculation of the flux growth rate, or a real change of the index with the total flux.

For six large active regions studied by \cite{Khlystova2013}, the flux growth rate chaotically varies in a range of (0.4 -1.0)$\times$10\textsuperscript{20} Mx hour\textsuperscript{-1} when the total flux varies in a range of (0.8-2)$\times$10\textsuperscript{22} Mx. The reported flux growth rate is lower than our estimates for the same values of the flux, see Figure \ref{Fig6}. The reason might be again the calculation method: in \cite{Khlystova2013}, as well as in \cite{Otsuji2011}, the flux growth rate was derived by dividing the saturated flux over the entire emergence time, whereas in our study, the time derivative $R(t)$ of the total flux was used to calculate the maximal flux growth rate.

Scrutinized inspection of movies of emergence allows us to detect a typical difference between the two emergence types. The "rapid"-emergence scenario is consistent with a "two-step" emergence of a single twisted flux tube \citep{Archontis2004, Toriumi2010, Toriumi2011} and with magnetic flux "shredding" described in \cite{Cheung2010, Cheung2014}. Frequently, for these active regions, a rapid expansion of an oval-shaped pattern of mixed polarity magnetic threads is observed. The leading and following sunspots are formed later, approximately near the peak of the flux growth rate, t\textsuperscript{*} time in Figure \ref{Fig7}. Before the t\textsuperscript{*}-time, the flattened and horizontally expanded (during the first step of sub-photospheric rising) flux tube blows up into the photosphere; by the t\textsuperscript{*}-time, the cross-section of the tube already rose above the photosphere, and later the leading and following footpoints of the tube become more vertical; the flux growth rate diminishes. It is possible to suggest, at least, two reasons for the decrease of $R(t)$ at this stage of emergence. First, as the apex of the tube rises, the contribution to the total flux from the serpentine field lines diminishes due to cancellation, magnetic diffusion between opposite polarities, retraction of U-loops, \textit{etc.}  \citep{Cheung2008, Centeno2012}. Second, the azimuth component $B$\textsuperscript{$\theta$} of the magnetic field of the tube predominantly contributes to the line-of-sight flux at the very early stage of emergence, when the tube is nearly horizontal. As the apex rises and the footpoints become more vertical, $B$\textsuperscript{$\theta$} contributes to the transverse flux \citep{Cheung2007}.

The events of the "gradual" emergence can then be explained as a consecutive rising of several flux tubes at the same location, in a full agreement with the suggestions by  \cite{Zwaan1985} and \cite{Toriumi2014}. Each tube is then experiencing the "two-step" emergence process: indeed, bubbles of mixed polarity serpentine-line features always accompany the consecutive emergence of dipoles of any size.

A question why the "rapid"-emergence active regions display the higher normalized flux emergence rate (in other words, they are able to experience the more intensive emergence) than the "gradually" emerging active regions, is still open.

A possible explanation can be offered that the sub-photospheric twisting motions of the roots of emerging flux tube intensify during the emergence and thus generate additional azimuthal component of the magnetic field. The effect can be more pronounced in the emergence of a single strong flux tube than in an ensemble of rather weak flux tubes. In the more general way, the sub-photospheric turbulent dynamo action (see, \textit{e.g.},  \cite{Brandenburg2012, Sokoloff2015}) can manifest itself in this way, at the emergence stage of magnetic structures.

Another explanation can be suggested as well. We see that "rapid"-emergence ARs usually exhibit emergence of a flattened single bundle, while "gradual"-emergence ARs are formed by gradual, extended in time appearance of several dipoles. This observation can be explained by different spatial orientation of magnetic flux bundle before the start of emergence. A horizontally-oriented flattened subsurface bundle would produce "rapid" emergence. Correspondingly, vertically-oriented nearly-flat bundle would result in consecutive appearance of the dipoles, one-by-one, as they rise through the photosphere. If this explanation is correct it raises another question -- why different ARs are oriented in different way in the upper layer of the photosphere during their pre-emergence stage.

Anyway, a further comparison with simulated data are needed to further clarify the one of the most intriguing phenomenon on the Sun -- the birth of an active region.



%

%

%
 \begin{acks}
 	
 	
SDO is a mission for NASA’s Living With a Star (LWS) program. The SDO/HMI data were provided by the Joint Science Operation Center (JSOC). The reported study was supported by the RFBR research projects 16-02-00221 A and 16-42-910493 and the Presidium of the Russian Academy of Science Program 7. VYu acknowledges support from AFOSR FA9550-15-1-0322 and NSF AGS-1250818 grants and Korea Astronomy and Space Science Institute. We thank International Space Science Institute for enabling interesting discussions. We are thankful to anonymous referee for critical comments and suggestions that helped very much to improve the paper.
 \end{acks}

%
%

\begin{thebibliography}{}


\bibitem[Archontis \emph{et al.}(2004)]
	{Archontis2004}
	Archontis, V., Moreno-Insertis, F., Galsgaard, K., Hood, A., O'Shea, E.: 2004, {\it Astron. Astroph.} {\bf 426}, 1047. \href{http://dx.doi.org/10.1051/0004-6361:20035934}{DOI} \href{http://adsabs.harvard.edu/abs/2004A\%26A...426.1047A}{ADS}

\bibitem[Babcock(1961)]
	{Babcock1961}
	Babcock, H.W.: 1961, {\it Astrophys. J.} {\bf 133}, 572. \href{http://dx.doi.org/10.1086/147060}{DOI} \href{http://adsabs.harvard.edu/abs/1961ApJ...133..572B}{ADS}

\bibitem[Brandenburg, Sokoloff, and Subramanian(2012)]
	{Brandenburg2012}
	Brandenburg, A., Sokoloff, D., Subramanian, K.: 2012, {\it Space Science Reviews} {\bf 169}, 123. \href{http://dx.doi.org/10.1007/s11214-012-9909-x}{DOI} \href{http://adsabs.harvard.edu/abs/2012SSRv..169..123B}{ADS}

\bibitem[Centeno(2012)]
	{Centeno2012}
	Centeno, R.: 2012, {\it Astrophys. J.} {\bf 759}, 72.  \href{http://dx.doi.org/10.1088/0004-637X/759/1/72}{DOI} \href{http://adsabs.harvard.edu/abs/2012ApJ...759...72C}{ADS}

\bibitem[Cheung and Isobe(2014)]
	{Cheung2014}
	Cheung, M.C.M., Isobe, H.: 2014, {\it Living Reviews in Solar Physics} {\bf 11}. \href{http://dx.doi.org/10.12942/lrsp-2014-3}{DOI} \href{http://adsabs.harvard.edu/abs/2014LRSP...11....3C}{ADS}

\bibitem[Cheung \emph{et al.}(2010)]
	{Cheung2010}
	Cheung, M.C.M., Rempel, M., Title, A.M., Sch{\"u}ssler, M.: 2010, {\it Astrophys. J.} {\bf 720}, 233. \href{http://dx.doi.org/10.1088/0004-637X/720/1/233}{DOI} \href{http://adsabs.harvard.edu/abs/2010ApJ...720..233C}{ADS}


\bibitem[Cheung, Sch{\"u}ssler, and Moreno-Insertis(2007)]
	{Cheung2007}
	Cheung, M.C.M., Sch{\"u}ssler, M., Moreno-Insertis, F.: 2007, {\it Astron. Astroph.} {\bf 467}, 703. \href{http://dx.doi.org/10.1051/0004-6361:20077048}{DOI} \href{http://adsabs.harvard.edu/abs/2007A\%26A...467..703C}{ADS}

\bibitem[Cheung \emph{et al.}(2008)]
	{Cheung2008}
	Cheung, M.C.M., Sch{\"u}ssler, M., Tarbell, T.D., Title, A.M.: 2008, {\it Astrophys. J.} {\bf 687}, 1373-1387. \href{http://dx.doi.org/10.1086/591245}{DOI} \href{http://adsabs.harvard.edu/abs/2008ApJ...687.1373C}{ADS}

\bibitem[Emonet and Moreno-Insertis(1998)]
	{Emonet1998}
	Emonet, T., Moreno-Insertis, F.: 1998, {\it Astrophys. J.} {\bf 492}, 804. \href{http://dx.doi.org/10.1086/305074}{DOI} \href{http://adsabs.harvard.edu/abs/1998ApJ...492..804E}{ADS}

\bibitem[Jouve and Brun(2009)]
	{Jouve2009}
	Jouve, L., Brun, A.S.: 2009, {\it Astrophys. J.} {\bf 701}, 1300. \href{http://dx.doi.org/10.1088/0004-637X/701/2/1300}{DOI} \href{http://adsabs.harvard.edu/abs/2009ApJ...701.1300J}{ADS}

\bibitem[Fan(2009)]
	{Fan2009}
	Fan, Y.: 2009, {\it Living Reviews in Solar Physics} {\bf 6}, 4. \href{http://dx.doi.org/10.12942/lrsp-2009-4}{DOI} \href{http://adsabs.harvard.edu/abs/2009LRSP....6....4F}{ADS}

\bibitem
	[Fu and Welsch(2016)]
	{Fu2016}Fu, Y., Welsch, B.T.: 2016, {\it Solar Phys.} {\bf 291}, 383. \href{http://dx.doi.org/10.1007/s11207-016-0851-z}{DOI} \href{http://adsabs.harvard.edu/abs/2016SoPh..291..383F}{ADS}

\bibitem[Hagenaar(2001)]
	{Hagenaar2001}
	Hagenaar, H.J.: 2001, {\it Astrophys. J.} {\bf 555}, 448. \href{http://dx.doi.org/10.1086/321448}{DOI} \href{http://adsabs.harvard.edu/abs/2001ApJ...555..448H}{ADS}


\bibitem[Khlystova(2013)]
	{Khlystova2013}
	Khlystova, A.: 2013, {\it Solar Phys.} {\bf 284}, 329. \href{http://dx.doi.org/10.1007/s11207-012-0193-4}{DOI} \href{http://adsabs.harvard.edu/abs/2013SoPh..284..329K}{ADS}
	
\bibitem[Kutsenko and Abramenko(2016)]
	{Kutsenko2016}
	Kutsenko, A.S., Abramenko, V.I.: 2016, {\it Solar Phys.}. \href{http://dx.doi.org/10.1007/s11207-016-0940-z}{DOI} \href{http://adsabs.harvard.edu/abs/2016SoPh..tmp..104K}{ADS}

\bibitem[Leighton(1969)]
	{Leighton1969}
	Leighton, R.B.: 1969, {\it Astrophys. J.} {\bf 156}, 1. \href{http://dx.doi.org/10.1086/149943}{DOI} \href{http://adsabs.harvard.edu/abs/1969ApJ...156....1L}{ADS}

\bibitem[Liu \emph{et al.}(2012)]
	{Liu2012}
	Liu, Y., Hoeksema, J.T., Scherrer, P.H., Schou, J., Couvidat, S., Bush, R.I., Duvall, T.L., Hayashi, K., Sun, X., Zhao, X.: 2012, {\it Solar Phys.} {\bf 279}, 295. \href{http://dx.doi.org/10.1007/s11207-012-9976-x}{DOI} \href{http://adsabs.harvard.edu/abs/2012SoPh..279..295L}{ADS} 

\bibitem[Moreno-Insertis and Emonet(1996)]
	{MorenoInsertis1996}
	Moreno-Insertis, F., Emonet, T.: 1996, {\it Astrophys. J.} {\bf 472}, L53. \href{http://dx.doi.org/10.1086/310360}{DOI} \href{http://adsabs.harvard.edu/abs/1996ApJ...472L..53M}{ADS}

\bibitem[Otsuji \emph{et al.}(2011)]
	{Otsuji2011}
	Otsuji, K., Kitai, R., Ichimoto, K., Shibata, K.: 2011, {\it Pub. Astron. Soc. Japan} {\bf 63}, 1047. \href{http://dx.doi.org/10.1093/pasj/63.5.1047}{DOI} \href{http://adsabs.harvard.edu/abs/2011PASJ...63.1047O}{ADS}

\bibitem[Parker(1975)]
	{Parker1975}
	Parker, E.N.: 1975, {\it Astrophys. J.} {\bf 198}, 205.	\href{http://dx.doi.org/10.1086/153593}{DOI} \href{http://adsabs.harvard.edu/abs/1975ApJ...198..205P}{ADS}

\bibitem[Pesnell, Thompson, and Chamberlin(2012)]
	{Pesnell2012}
	Pesnell, W.D., Thompson, B.J., Chamberlin, P.C.: 2012, {\it Solar Phys.} {\bf 275}, 3. \href{http://dx.doi.org/10.1007/s11207-011-9841-3}{DOI} \href{http://adsabs.harvard.edu/abs/2012SoPh..275....3P}{ADS} 

\bibitem[Rempel and Cheung(2014)]
	{Rempel2014}
	Rempel, M., Cheung, M.C.M.: 2014, {\it Astrophys. J.} {\bf 785}, 90. \href{http://dx.doi.org/10.1088/0004-637X/785/2/90}{DOI} \href{http://adsabs.harvard.edu/abs/2014ApJ...785...90R}{ADS}

\bibitem[Scherrer \emph{et al.}(2012)]
	{Scherrer2012}
	Scherrer, P.H., Schou, J., Bush, R.I., Kosovichev, A.G., Bogart, R.S., Hoeksema, J.T., Liu, Y., Duvall, T.L., Zhao, J., Title, A.M., Schrijver, C.J., Tarbell, T.D., Tomczyk, S.: 2012, {\it Solar Phys.} {\bf 275}, 207. \href{http://dx.doi.org/10.1007/s11207-011-9834-2}{DOI} \href{http://adsabs.harvard.edu/abs/2012SoPh..275..207S}{ADS}

\bibitem[Schou \emph{et al.}(2012)]
	{Schou2012}
	Schou, J., Scherrer, P.H., Bush, R.I., Wachter, R., Couvidat, S., Rabello-Soares, M.C., Bogart, R.S., Hoeksema, J.T., Liu, Y., Duvall, T.L., Akin, D.J., Allard, B.A., Miles, J.W., Rairden, R., Shine, R.A., Tarbell, T.D., Title, A.M., Wolfson, C.J., Elmore, D.F., Norton, A.A., Tomczyk, S.: 2012, {\it Solar Phys.} {\bf 275}, 229. \href{http://dx.doi.org/10.1007/s11207-011-9842-2}{DOI} \href{http://adsabs.harvard.edu/abs/2012SoPh..275..229S}{ADS}

\bibitem[Schuessler(1979)]
	{Schuessler1979}
	Schuessler, M.: 1979, {\it Astron. Astroph.} {\bf 71}, 79. \href{http://adsabs.harvard.edu/abs/1979A\%26A....71...79S}{ADS}

\bibitem[Smirnova \emph{et al.}(2013)]
	{Smirnova2013}
	Smirnova, V., Efremov, V.I., Parfinenko, L.D., Riehokainen, A., Solov'ev, A.A.: 2013, {\it Astron. Astroph.} {\bf 554}, A121. \href{http://dx.doi.org/10.1051/0004-6361/201220825}{DOI} \href{http://adsabs.harvard.edu/abs/2013A\%26A...554A.121S}{ADS} 

\bibitem[Sokoloff, Khlystova, and Abramenko(2015)]
	{Sokoloff2015}
	Sokoloff, D., Khlystova, A., Abramenko, V.: 2015, {\it Mon. Not. Roy. Astron. Soc.} {\bf 451}, 1522. \href{http://dx.doi.org/10.1093/mnras/stv1036}{DOI} \href{http://adsabs.harvard.edu/abs/2015MNRAS.451.1522S}{ADS}

\bibitem[Spruit(1981)]
	{Spruit1981}
	Spruit, H.C.: 1981, {\it Astron. Astroph.} {\bf 98}, 155. \href{http://adsabs.harvard.edu/abs/1981A\%26A....98..155S}{ADS}

\bibitem[Toriumi, Hayashi, and Yokoyama(2014)]
	{Toriumi2014}
	Toriumi, S., Hayashi, K., Yokoyama, T.: 2014, {\it Astrophys. J.} {\bf 794}, 19. \href{http://dx.doi.org/10.1088/0004-637X/794/1/19}{DOI} \href{http://adsabs.harvard.edu/abs/2014ApJ...794...19T}{ADS}

\bibitem[Toriumi and Yokoyama(2010)]
	{Toriumi2010}
	Toriumi, S., Yokoyama, T.: 2010, {\it Astrophys. J.} {\bf 714}, 505. \href{http://dx.doi.org/10.1088/0004-637X/714/1/505}{DOI} \href{http://adsabs.harvard.edu/abs/2010ApJ...714..505T}{ADS}

\bibitem[Toriumi and Yokoyama(2011)]
	{Toriumi2011}
	Toriumi, S., Yokoyama, T.: 2011, {\it Astrophys. J.} {\bf 735}, 126. \href{http://dx.doi.org/10.1088/0004-637X/735/2/126}{DOI} \href{http://adsabs.harvard.edu/abs/2011ApJ...735..126T}{ADS}

\bibitem[van Driel-Gesztelyi \emph{et al.}(2003)]
	{vanDrielGesztelyi2003}
	van Driel-Gesztelyi, L., D{\'e}moulin, P., Mandrini, C.H., Harra, L., Klimchuk, J.A.: 2003, {\it Astrophys. J.} {\bf 586}, 579. \href{http://dx.doi.org/10.1086/367633}{DOI} \href{http://adsabs.harvard.edu/abs/2003ApJ...586..579V}{ADS}

\bibitem[Zwaan(1985)]
	{Zwaan1985}
	Zwaan, C.: 1985, {\it Solar Phys.} {\bf 100}, 397. \href{http://dx.doi.org/10.1007/BF00158438}{DOI} \href{http://adsabs.harvard.edu/abs/1985SoPh..100..397Z}{ADS}

 \end{thebibliography}
%

\end{article} 
\end{document}